\documentclass[11pt]{article}
\usepackage{mathrsfs}
\usepackage{amsfonts}
\usepackage{amssymb}
\usepackage{amsmath}
\usepackage{graphicx}
\usepackage{cite}
\usepackage{bm}

\usepackage{mathrsfs}
\usepackage{amsfonts}
\usepackage{amssymb}
\usepackage{amsmath}
\usepackage{graphicx}
\usepackage{cite}
\usepackage{bm}

\newtheorem{Th}{Theorem}[section]
 \newtheorem{Lem}{Lemma}[section]
\newtheorem{Cor}{Corollary}[section]

\newtheorem{defn}{Definition}[section]
\newtheorem{Rem}{Remark}[section]
\newtheorem{Exa}{Example}[section]
\newtheorem{Prob}{Problem}[section]
\newtheorem{Method}{Method}[section]

\def\R{ \mathbb{R}}

 \def\ur{\underline{p}}
 \def\uv{\underline{m}}

\def\i{{\bf i}}

 \def\j{{\bf j}}
 \def\k{{\bf k}}

 \def\uQ{\underline{Q}}

\date{}
\begin{document}
%\pagestyle{plain}
%\begin{frontmatter}
\pagestyle{plain}
\pagenumbering{arabic}

\title{\bf Quaternion Hardy Functions for Local Images Feature Representation}
%Edge Detection base on the Quaternion Analytic function}

\author{Xiaoxiao Hu  \thanks{
The First Affiliated Hospital of Wenzhou Medical University
, Wenzhou, Zhejiang, China. Email: huxiaoxiao3650@163.com.}, Kit-Ian Kou \  \thanks{Corresponding author: Kit-Ian Kou. Department of Mathematics, University of Macau, Macao (Via Hong
Kong). Email: kikou@umac.mo,}
Yangxing Ting\thanks{ Institute of Digitized Medicine, Wenzhou Medical University
, Wenzhou, Zhejiang, China. Email: 1281961491@qq.com.}
}

\maketitle

%%%%%%%%%%%%%%%%%%%%%%%%%%%%%%%%%%%%%%%%%%%%%%%%%%%%%%%%%
\begin{abstract}
This paper is concerned with the applications of local features of the quaternion Hardy function. The feature information can be provided by the polar form of the quaternion Hardy function, such as the local attenuation and local phase vector. By using the generalized Cauchy-Riemann equations of local features, four various kinds of edge detectors are newly developed.  The experiment results show that the amplitude-based edge detectors perform the envelope of test images while the phase-based edge detectors give the detail of the test images. Our proposed algorithms greatly outperform existing schemes.
\end{abstract}

%\subjclass{Primary 44A15; 70G10; 35105}
{\bf Keywords}:
Quaternion;   quaternion analytic signal; edge detection;   Poisson operator.
\bigskip

{\bf AMS Mathematical Subject Classification:}  44A15;  35F15; 70G10
%\end{keyword}
%%\end{frontmatter}

%==================================================================================
\section{Introduction}\label{hxx1}
 The edge detection is a  preprocessing step  in image analysis and computer vision. It aims to preserve the important structural properties in an image, hence, it is  an  crucial part in the  computer vision-based applications. Such as color edge detection \cite{li2019Noise}, motion \cite{jain1995machine},  heart rate detection \cite{wang2020Ballistocardiogram},
% image enhancement \cite{sun2002triangle},
  and so on \cite{sonka2014image,zuppinger2017edge}.  There are a variety of edge detectors which have been devised \cite{li2015survey}. The classical differentiation based edge detection identifies pixel locations with brightness changes by gradient operators, such as the Canny edge detection which is the popular choice for applications.  The disadvantage of the differentiation based edge detection methods is the noise-sensitivity. In our paper, our approaches are based on the quaternion Hardy function, which is obtained by quaternion analytic signal  being holomorphically extended into the upper (lower) half-space of two quaternion  variables.
  \par
 The analytic signal has lots of applications in signal and image processing \cite{Gengel2019phase,hahn1996hilbert,Venkitaraman2019Hilbert}. The polar form of an analytic signal provides the signal features, such as the local amplitude and  phase, which contain  structural information of the signal,  this fact  enlightens  us to apply them on image edge detection.
  By an original signal convoluting with the Poisson and conjugate Poisson integral, respectively, the analytic signal can be analytic  continued  into  the  upper (lower) Hardy space to be a Hardy  function.
Therefore Hardy function can handle various types of noises. Motivated by their excellent properties, high dimension extension was studied in several complex variable space \cite{hahn1996hilbert,hahn2005wigner}, quaternion space \cite{felsberg2004monogenic,felsberg2001monogenic,bernstein2013generalized} and Clifford space \cite{kou2002paley,yang2017edge}.  Generalizations under various transformations were analyzed, such as quaternion Fourier transform(QFT) \cite{bulow2001hypercomplex,bernstein2013generalized,hahn2005wigner} and quaternion linear canonical transform\cite{kou2016envelope}.

By convoluting Riesz kernel in higher dimensions,
M.Felsberg and G.Sommer in 2001 \cite{felsberg2001monogenic} first studied the monogenic signal. Later,  in 2004, they \cite{felsberg2004monogenic} proposed the monogenic scale-space, which  is  extension of the Hardy space in the higher dimension.
The relationship between the local features of the monogenic scale function in the intrinsically 1D cases are deduced  in \cite{felsberg2004monogenic}, they satisfied the  Cauchy-Riemann equations. In this paper, they also
 proposed a edge detection approach based on  the phase of  {\it intrinsically 1D monogenic signals}, it is called the  \textbf{differential phase congruency (DPC)} method.
While in the case of the function  is not intrinsically 1D, the study was derived by Yan et al \cite{yang2017edge}, The  \textbf{ modified differential phase congruency (MDPC)} was analyzed as the edge detector in their paper \cite{yang2017edge}.

In \cite{bernstein2013generalized},the quaternion analytic signal is  one of the  generalization of the analytic signal in high dimension.
It corresponds to a boundary value of a quaternion  Hardy function  which is holomorphic in two quaternion variables $ s_1 =t_1+\i y_1, s_2=t_2+\j y_2$. The quaternion Hardy function also provides the local features: the local attenuation, local phase and the local phase-vector. To the best of our knowledge, the relations between the features of  quaternion Hardy function have not been carried out. In this paper, we will study the connection between them. The contributions of this paper are summarized as follows.

\begin{itemize}
  \item  By the generalized Cauchy-Riemann equations, the relations between the features  of quaternion Hardy function are derived.

  \item  The Phase-Based and Amplitude-based methods for edge detection filters are proposed. Theoretical and experiment results are established, respectively.
\end{itemize}
 The remainder of the paper is organized  as follows.
 Section \ref{hxx2} reviews some basic concepts.  Section \ref{hxx3} gives the relationship of the features of quaternion Hardy function and proposes four novel  edge detection  approaches   which are  based on the features of quaternion Hardy function. Section \ref{hxx5} shows the experiment results. Section \ref{hxx6} gives the conclusion and future work.

%----------------------------------------------------------------------
\section{Preliminaries}\label{hxx2}
This section is devoted to the exposition of basic preliminary  material which we use extensively throughout of this paper.

\subsection{Quaternion Algebra}
The notation   $\mathbb{H}$ means the {\it Hamiltonian skew field of quaternions}, which
%are  isomorphic to $  {\it Cl}_{0, 2},$
% $\mathbb{H} \simeq {\it Cl}_{0, 2}=span\{1,{\bf e}_1, {\bf e}_2, {\bf e}_1{\bf e}_2   \}  $,
can represent   multidimensional signal as a holistic signal \cite{zou2016quaternion, xia2018quaternion,Safarian2019quaternion,Jin2019effective,liu2019constrained, liu2018stability}.
 A quaternion-value number takes a form
\begin{eqnarray}\label{u1}
Q:=Q_0+{\i}Q_1+{\j}Q_2+{\k}Q_3,
\end{eqnarray}
where $ Q_k\in\mathbb{R}, k=0,1,2,3,$  $\{1, \i, \j, \k \}$ is a quaternion orthogonal basis, and obeys the Hamilton's multiplication rules:
\begin{eqnarray*}
 \i^2=  \j^2= \k^2= \i\j\k=-1.
\end{eqnarray*} Denote the various notations of quaternion-value number $ Q$ as follows
\begin{itemize}
  \item the  Scalar part : ${\rm{Sc}}[Q]:=Q_0$,
  \item the $\i$   part : ${\rm{Vec}(\i)}[Q]:=Q_1$,
  \item the  $\j$  part : ${\rm{Vec}(\j)}[Q]:=Q_2$,
  \item the  $\k$  part : ${\rm{Vec}(\k)}[Q]:=Q_3$,
  \item the Vector part: $ \underline{Q}$:=${\bf i}Q_1+{\bf j}Q_2+{\bf k}Q_3.$
\end{itemize}

$
Q^{\ast} :=Q_0-{\bf i}Q_1-{\bf j}Q_2-{\bf k}Q_3.
$
is denoted the {\it conjugate} of a quaternion $ Q$.
$
|Q| := \sqrt{QQ^{\ast}} = \sqrt{Q^{\ast}Q} = \sqrt{Q_0^2+Q_1^2+Q_2^2+Q_3^2}.
$
 is the modulus of $Q\in\mathbb{H}$.
It is
easy to verify that for $0 \not= Q_0+\uQ\in \mathbb{H}$, the inverse of $Q$ can be defined by
$$Q=(Q_0+\uQ)^{-1} := \frac{\overline{Q_0+\uQ}}{|QQ_0+\uQ|^2}.$$

From Eq.$(\ref{u1})$, it follows that a quaternion-value  function $f:\mathbb{R}^2\to\mathbb{H}$ can be expressed as
\begin{eqnarray*}
g(t_1, t_2)=g_0(t_1, t_2)+\i g_1(t_1, t_2)+\j g_2(t_1, t_2)+ \k g_3(t_1, t_2),
\end{eqnarray*}
where $f_{n}\in\mathbb{R}, n=0,1,2,3$.
Let $L^{p}(\mathbb{R}^2 , \mathbb{H}),$ ( integers $ p \geq$ 1) be  the space of all quaternion-value function in $\mathbb{R}^2,$ whose quaternion-value modules are
defined by
\begin{eqnarray*}
L^{p}(\mathbb{R}^2, \mathbb{H}):=\{ g| g:\mathbb{R}^2 \to \mathbb{H}, \Vert g\Vert_{p}:=\left (\int_{\mathbb{R}^2} |g(t_{1},t_{2})|^p dt_{1}dt_{2} \right )^{\frac{1}{p}} <\infty \}.
\end{eqnarray*}

There are different  kinds of Quaternion Fourier transform (QFT) \cite{ell2014quaternion,hu2016quaternion}. In this paper, we derive the main results associated with the following   QFT\cite{ell2014quaternion,hu2016quaternion} , which is defined by
 \begin{eqnarray*}
\mathcal{F}[g](\omega_1, \omega_2):=\int_{\mathbb{R}^2} e^{-{ \i}\omega_1 t_1}g(t_1, t_2)e^{{- \j} \omega_2 t_2}dt_1dt_2.\label{hu24}
\end{eqnarray*}
where  $g\in L^{1}(\mathbb{R}^2, \mathbb{H}).$
%The QFT is   invertible and its inverse is  expressed as \cite{hu2016quaternion}:
%Set $f$ and $\mathcal{F}[f] \in L^{1}(\mathbb{R}^2, \mathbb{H}),  $
%\begin{eqnarray}\label{IFT}
%f(x,y)=\frac{1}{4\pi^{2}}\int_{\mathbb{R}^2} e^{\i \omega_1 x}\mathcal{F}[f](\omega_1, \omega_2)e^{\j \omega_2 y}dudv.
%\end{eqnarray}
%

%------------------------------------------------

%---------------------------------------------------

\subsection{Quaternion Hardy scale Space}
There are various ways to analysis analytic signal to higher dimensional space \cite{bernstein2013generalized, yang2012phase}. In this paper, the generalization is motivated by \cite{bernstein2013generalized} using the quaternion analysis. Firstly, let's recall
 definitions and properties  of the  quaternion partial and total Hilbert transforms associated with QFT \cite{hu2017phase,kou2016envelope} .

\begin{defn} \label{def1} \cite{kou2016envelope}
The quaternion  partial Hilbert transform  QPHT $\mathcal{H}_{1}$
%of$f$ along the $x_1$-axis or  $x_2$-axis,
and the quaternion  total Hilbert transform  QTHT $\mathcal{H}_{2} $
%along the $x$, $y$ axes
of $ f$ are given by
\begin{eqnarray}
\mathcal{H}_{1}[g(\cdot,t_2)](t_1)&:=&\frac{1}{\pi}p.v.\int_{\mathbb{R}}\frac{g(\tau,t_2)}{t_1-\tau}d\tau, \label{deH1}\\
\mathcal{H}_{1}[g(t_1,\cdot)](t_2)&:=&\frac{1}{\pi}p.v.\int_{\mathbb{R}}\frac{g(t_1,\tau)}{t_2-\tau}d\tau,\label{deH2}\\
\mathcal{H}_{2}[g(\cdot,\cdot)](t_1,t_2)&:=&\frac{1}{\pi^{2}}p.v.\int_{\mathbb{R}^{2}}\frac{g(\tau,\nu)}{(t_1-\tau)(t_2-\nu)}d\tau d\nu,\label{H3}
\end{eqnarray}
where $f$  is a generally quaternion-value function such that Eqs. $(\ref{deH1}), (\ref{deH2})$ and $ (\ref{H3})$ are well defined.
\end{defn}

%%=============================================================
%\begin{Lem}\cite{kou2016envelope} \label{invle}
%  Suppose  $f\in L^{2}(\mathbb{R}^{2}, \mathbb{H})$, then
%\begin{eqnarray*}
% \mathcal{F}[\mathcal{H}_{1}[f(\cdot,t_2)](x_1)]( \omega_{1},\omega_{2})&=&-sgn(\omega_{1})\i \mathcal{F}[f](\omega_{1},\omega_{2}).\label{F1}\\
% \mathcal{F}[\mathcal{H}_{1}[f(t_1,\cdot)](t_2)](\omega_{1},\omega_{2})&=&-sgn(\omega_{2}) \mathcal{F}[f](\omega_{1},\omega_{2})\j.\label{F2}\\
% \mathcal{F}[\mathcal{H}_{2}[f]](\omega_{1},\omega_{2})&=&sgn(\omega_{1}) sgn(\omega_{2})\i\mathcal{F}[f](\omega_{1},\omega_{2})\j.\nonumber\label{F3}
% \end{eqnarray*}
%\end{Lem}
%%=========================================================
%\begin{Lem}\cite{hu2016quaternion} \label{edge}
%If $f \in L^{1}(\mathbb{R}^{2}, \mathbb{H})$ has derivatives in the $ L^{1}(\mathbb{R}^{2}, \mathbb{H})$ norm of all orders $ \leq m+n$, then
%\begin{eqnarray*}
%\mathcal{F}\left[\frac{\partial ^{m+n}}{\partial t_1^{m}\partial t_2^{n}}f\right ] (\omega_{1},\omega_{2})=(\i \omega_{1})^{m}\mathcal{F}[f](\omega_{1},\omega_{2})(\j \omega_{2})^{n},
%\end{eqnarray*}
%\end{Lem}
%================================================
Taking the QFT of Eqs. $(\ref{deH1}), (\ref{deH2})$ and $ (\ref{H3})$ , we obtain that
\begin{eqnarray*}
\mathcal{F}[\mathcal{H}_{1}[g(\cdot,t_2)](t_1)](\omega_{1},\omega_{2})&=&-\frac{1}{|\omega_{1} |} \mathcal{ F}[\partial g/ \partial t_1](\omega_{1},\omega_{2}).\\
\mathcal{F}[\mathcal{H}_{1}[g(t_1,\cdot)](t_2)](\omega_{1},\omega_{2})&=&-\frac{1}{|\omega_{2} |} \mathcal{ F}[\partial g /\partial t_2](\omega_{1},\omega_{2}).\\
\mathcal{F}[\mathcal{H}_{2}[g]](\omega_{1},\omega_{2})&=&-\frac{1}{|\omega_{1} \omega_{2}|} \mathcal{ F}[\partial ^{2}g/\partial t_1\partial t_2](\omega_{1},\omega_{2}).\\
\end{eqnarray*}

Hence, the QPHT $\mathcal{H}_{i}$ ($i=1, 2$) and QTHT $\mathcal{H}_{2}$ have functions of  the edge detectors \cite{hu2016quaternion}.

The quaternion analytic  signal \cite{hu2017phase} was defined by an
original signal and its QPHT and QTHT as following.

%--------------------------------------------------------
\begin{defn}[Quaternion Analytic Signal]\label{qftAD}\cite{hu2017phase}
 The quaternion analytic signal $g_{q}$ of $g$ can be defined by
\begin{eqnarray*}
g_{q}(t_1,t_2):=g(t_1,t_2)+\i\mathcal{H}_{1}[g(\cdot,t_2)](t_1)+\mathcal{H}_{1}[g(t_1,\cdot)](t_2)\j+\i\mathcal{H}_{2}[g](t_1,t_2)\j,
\end{eqnarray*}
where $g(t_1,t_2)$ is a quaternion-value function such  that $ g_{q}$ is well defined, that is, Eqs. $(\ref{deH1}), (\ref{deH2})$ and $ (\ref{H3})$ are well defined.
\end{defn}

%==============================
\begin{Rem}
  When $f \in L^{2}(\mathbb{R}^{2}, \mathbb{R}), $  then the definition is the same as in papers \cite{bulow2001hypercomplex, bernstein2013generalized}.
\end{Rem}
%----------------------------------
Let $ g\in L^{2}(\mathbb{R}^{2}, \mathbb{H})$,  we have
$$ \mathcal{F}[g_{q}](\omega_{1},\omega_{2})=(1+sgn(\omega_{1}))(1+sgn(\omega_{2})) \mathcal{F}[g](\omega_{1},\omega_{2}).$$
%----------------------------

Let us now study  quaternion Hardy scales space, which  is the Hardy space in quaternion algebra setting.
%%--------------------
%{\bf the Quaternion Hardy space is first time defined in this paper? Otherwise, please give the citation.}
%
%

%------------------------------------------

\begin{defn}\cite{hu2017phase}{\bf(Quaternion Hardy Scales Space $\mathbb{QS}(\mathbb{CS}_{\i \j}^{+}, \mathbb{H})$)}  \label{QHSS} Quaternion  Hardy scales  Space  $\mathbb{QS}(\mathbb{CS}, \mathbb{H})$ is the class of quaternion Hardy functions (QHF) $g(s_1, s_2)$ which are
defined on the upper half space
$ \mathbb{CS}$
and satisfies the  following conditions:
 \begin{enumerate}
   \item $\frac{\partial}{\partial \overline{s_{1}}}g=0,$
   \item $ g\frac{\partial}{\partial \overline{s_{2}}}=0, $
   \item $\int_{\mathbb{R}^2} |g(t_{1}+\i y_{1}, tx_{2}+\j y_{2})|^{2}dt_{1}dt_{2} < C,$ for $ y_{1}>0, y_{2}>0 $,
 \end{enumerate}
where  $ \mathbb{CS}:= \{( s_1, s_2)|s_1:=t_1+\i y_1, s_2:=t_2+\j y_2, y_1>0, y_2 > 0\}. $  \end{defn}

Due to the non-commutativity of the quaternions,  the operators  $\frac{\partial}{\partial \overline{s_{1}}} $  and$\frac{\partial}{\partial \overline{s_{2}}} $ are  applied  to  quaternion-value function from the left and right,respectively.  The parameters $y_1$ and $y_2$ are regarded as the scales.

%Quaternion Cauchy kernel function as an example of quaternion Hardy function.
%%--------------------------
%\begin{Exa}
% Let
%$$ E(z_1,z_2)={z_{1}^{\ast}z_{2}^{\ast} \over |z_{1}|^{2} |z_{2}|^{2}}$$ be the {\it Cauchy kernel} defined in $ \mathbb{H}\setminus
%\{0\}$. It is easy to verify that $E$ is a quaternion Hardy function in $\mathbb{H}\setminus
%\{0\}$ .
%\end{Exa}
%%------------------------------

Studying the Riemann-Hilbert problem in $ \mathbb{CS}, f\in L^{2}(\mathbb{R}^2, \mathbb{H}),$
\begin{eqnarray*}\label{z1}
\frac{\partial}{\partial \overline{z_{1}}}g( s_1, s_2)=0, \quad ( s_1, s_2) \in \mathbb{CS}.
\end{eqnarray*}
\begin{eqnarray*}\label{z2}
 g( s_1, s_2)\frac{\partial}{\partial \overline{s_{2}}}= 0, \quad ( s_1, s_2) \in \mathbb{CS}.
\end{eqnarray*}
\begin{eqnarray*}\label{z3}
\mathcal{F}[g](\omega_1,\omega_2)=\left(1+sign(\omega_1)\right)\left(1+sign(\omega_2)\right)\mathcal{F}[h](\omega_1,\omega_2), \quad (\omega_{1},\omega_{2}) \in \mathbb{R}^2,
\end{eqnarray*}
The solutions are obtained  by
\begin{eqnarray}\label{qhf}
g(s_{1},s_{2})=\frac{1}{2\pi \i} \int_{\mathbb{R}^{2}} \frac{4h(x,y)}{(x-s_{1})(y-s_{2})}dxdy\frac{1}{2\pi \j},
\end{eqnarray}
where $\frac{h(x,y)}{(x-s_{1})(y-s_{2})}= \frac{(\overline{(x-s_1)}h(x,y)\overline{(y-s_2))})} {|x-s_{1}|^{2} |y-s_{2}|^{2}},$
and hence from the Plemelj-Sokhotzkis formulae for quaternion \cite{hu2017phase},
\begin{eqnarray*}
g(t_{1},t_{2})=h(t_{1},t_{2})+\i\mathcal{H}_{1}[h(\cdot,t_2)](t_1)+\mathcal{H}_{1}[h(t_1,\cdot)](t_2)\j+\i\mathcal{H}_{2}[h](t_{1},t_{2})\j,
\end{eqnarray*}
 which tells us that the quaternion analytic signal $g$ corresponds to the boundary value of a quaternion Hardy function in $\mathbb{CS}.$
%-------------------------------------------------
\begin{Th}\cite{hu2017phase} \label{qqas}
A quaternion-value signal $f_{q}\in L^{2}(\mathbb{R}^{2}, \mathbb{H}),$ is a quaternion analytic signal of $f\in L^{2}(\mathbb{R}^{2}, \mathbb{H})  $ if and only if $f_{q}$ is boundary value of the quaternion Hardy function in $\mathbb{CS}.$
i.e..
There exits  quaternion Hardy function (QHF)   $f_{q}(s_{1},s_{2}) \in \mathbb{Q}(\mathbb{CS}, \mathbb{H}),$  such that
 $$f_{q}(s_{1},s_{2})=\frac{1}{2\pi \i} \int_{\mathbb{R}^{2}} \frac{4f(x,y)}{(x-s_{1})(y-s_{2})}dsdt\frac{1}{2\pi \j}=r+\i m_1 + m_2 \j +\i m_3 \j,$$
 and
\begin{eqnarray}\label{qaf}
f_{q}(t_{1}, t_{2})=\lim_{y_{1}\rightarrow 0^{+},y_{2}\rightarrow 0^{+}} f_{q}(t_1+\i y_1, t_2+\j y_2).
\end{eqnarray}
That is,
 \begin{eqnarray*}\label{qaf}
f(t_{1}, t_{2})=\lim_{y_{1}\rightarrow 0^{+},y_{2}\rightarrow 0^{+}} r(t_1+\i y_1; t_2+\i y_2),
\end{eqnarray*}
 \begin{eqnarray*}\label{qaf}
\mathcal{H}_{1}[f(\cdot,t_2)](t_1)=\lim_{y_{1}\rightarrow 0^{+},y_{2}\rightarrow 0^{+}}m_{1}(t_1+\i y_1; t_2+\i y_2),
\end{eqnarray*}
 \begin{eqnarray*}\label{qaf}
\mathcal{H}_{1}[f(t_1,\cdot)](t_2)=\lim_{y_{1}\rightarrow 0^{+},y_{2}\rightarrow 0^{+}}m_{2}(t_1+\i y_1; t_2+\i y_2),
\end{eqnarray*}
 \begin{eqnarray*}\label{qaf}
\mathcal{H}_{2}[f](t_1,t_2)=\lim_{y_{1}\rightarrow 0^{+},y_{2}\rightarrow 0^{+}}m_{3}(t_1+\i y_1; t_2+\i y_2),
\end{eqnarray*}
  where  the functions $r$ and $m_{n}, n=1,2,3$ are constructed by the Poisson $P_{y_{1}}(t_{1}): = \frac{y_1}{y_{1}^2+t_{1}^2}$and the conjugate Poisson $Q_{y_{1}}(t_{1}): = \frac{t_1}{y_{1}^2+t_{1}^2}$ integrals, respectively.  $ * $ denotes the 2D convolution operator of quaternion-value functions $ f$ and $g$ in  $\mathbb{R}^{2},$ i.e.,
 $  f*g(t_1,t_2)=\int_{\R^2} f(x,y)g(t_1-x,t_2-y)dxdy $.
That is,
\begin{eqnarray} \label{poisson}
r(t_1+\i y_1; t_2+\i y_2)&=&r*P_{y_{1}}(t_{1})P_{y_{2}}(t_{2})\nonumber\\
&=&\frac{1}{\pi^2}\int_{\R^2}\frac{y_1 y_2 r(x,y)}{(y_{1}^2+(t_{1}-x)^2)(y_{2}^2+(t_{2}-y)^2)}dxdy\nonumber \\
m_{1}(t_1+\i y_1; t_2+\i y_2)&=&r*Q_{y_{1}}(t_{1})P_{y_{2}}(t_{2})\nonumber\\
&=&\frac{1}{\pi^2}\int_{\R^2}\frac{x_1 y_2 r(x,y)}{(y_{1}^2+(t_{1}-x)^2)(y_{2}^2+(t_{2}-y)^2)}dxdy,\nonumber\\
m_{2}(t_1+\i y_1; t_2+\i y_2)&=&r*P_{y_{1}}(t_{1})Q_{y_{2}}(t_{2})\nonumber\\
&=&\frac{1}{\pi^2}\int_{\R^2}\frac{y_1 t_2 r(x,y)}{(y_{1}^2+(t_{1}-x)^2)(y_{2}^2+(t_{2}-y)^2)}dxdy,\nonumber\\
m_{3}(t_1+\i y_1; t_2+\i y_2)&=&r*Q_{y_{1}}(t_{1})Q_{y_{2}}(t_{2})\nonumber\\
&=&\frac{1}{\pi^2}\int_{\R^2}\frac{x_1 x_2 r(x,y)}{(y_{1}^2+(t_{1}-x)^2)(y_{2}^2+(t_{2}-y)^2)}dxdy.\nonumber\\
\end{eqnarray}
\end{Th}
%-------------------------------------------------
%%%%%%%%%%%%%%%%%%%%%%%%%%%%%%%%%%%%%%%%%%%

\subsection{Local Features}\label{S3}

In this paper,  we study the case under the quaternion sense, in which  the quaternion analytic signal $g(t_1,t_2)$  is defined by  a real signal $ r\in L^{2}(\mathbb{R}^{2},\mathbb{R} )$,
 \begin{eqnarray*}
g(t_1,t_2)=r(t_1,t_2)+\i\mathcal{H}_{1}[r(\cdot,t_2)](t_1)+\mathcal{H}_{1}[r(t_1,\cdot)](t_2)\j+\i\mathcal{H}_{2}[r](t_1,t_2)\j,
\end{eqnarray*}
 and by  Theorem \ref{qqas}, its quaternion Hardy function is obtained as follows:
 \begin{eqnarray*}
 g(s_1, s_2)=r(s_1, s_2)+\i m_{1}(s_1, s_2)+m_{2}(s_1, s_2)\j+\i m_{3}(s_1, s_2)\j.
  \end{eqnarray*}
And  find  applications  of the local features of  quaternion Hardy function in image processing.

%\end{Rem}

%Under certain conditions,  it is possible to write the quaternion Hardy  function   in polar coordinate.
% According to, we can introduce the following features.

%--------------------------------------
\begin{defn} \textbf{[Local Features]}\cite{hu2017phase}
Suppose that quaternion  Hardy function $g(s_1, s_2) =r(s_1, s_2)+ \underline{m}(s_1, s_2) \in \mathbb{QS}(\mathbb{CS}, \mathbb{H})$ has the polar representation
\begin{eqnarray*}\label{polar}
g(s_1, s_2) &=&A(s_1, s_2)e^{\frac{\underline{m}(s_1, s_2)}{|\underline{m}(s_1, s_2)|}\theta}
 =e^{a(s_1, s_2)+\underline{p}(s_1, s_2)},
\end{eqnarray*}
 where
\begin{eqnarray*}\label{LA}
A(s_1, s_2): =|f(s_1, s_2)| =\sqrt{r(s_1, s_2)^2+|\underline{m}(s_1, s_2)|^2}\end{eqnarray*}
is called the \textbf{ local amplitude}, the energetic information of $ r$ is contained in it.
\begin{eqnarray}\label{atten}
a(s_1, s_2) :=\ln A(s_1, s_2)=\frac{1}{2}\ln(r^2(s_1, s_2)+|\underline{m}|^2(s_1, s_2))
\end{eqnarray} is called the \textbf{ local attenuation},
\begin{eqnarray}\label{PA}
\theta(s_1, s_2) :=\arctan \left(\frac{|\underline{m}(s_1, s_2)|}{r(s_1, s_2)}\right)\end{eqnarray}
is called the \textbf{ local phase}, the value of which is  is between $0$ and $\pi$.  The  structural information  of $r(s_1, s_2) $  is included in it.
%And  it also  describe how much a vector or quaternion
%number diverge from the real axis.

\begin{eqnarray}\label{lpv}
\underline{p}(s_1, s_2) :=\frac{\underline{m}(s_1, s_2)}{|\underline{m}(s_1, s_2)|}\theta(s_1, s_2),
\end{eqnarray}
 is called the \textbf{ local phase vector}.

%If $ \frac{\underline{v}(z_1, z_2)}{|\underline{v}(z_1, z_2)|} $ is a constant vector $ \underline{\mathbf{n}},$
% then  $f$ is called \textbf{ intrinsically 1D  }, let  $ \underline{p}_{in1}: = \underline{n}\theta (z_1, z_2)$.

 %The \textbf{   partial instantaneous frequencies } are defined by
%$\frac{\partial \theta}{\partial x_1}$ and $\frac{\partial \theta}{\partial x_2}.$
%    \item The \textbf{ partial scale  derivatives of  $\theta$} are defined by
%$\frac{\partial \theta}{\partial y_1}$ and $\frac{\partial \theta}{\partial y_2}.$

\end{defn}
We are now ready to proceed the main results.

%%-----------------------------------------
\section{Phase-Based and Amplitude-based approaches} \label{hxx3}

\subsection {Theoretical Basis  }

Suppose a complex function $C(t_1,t_2)= r(t_1,t_2)+\i m(t_1,t_2)=e^{a(t_1,t_2)+\i \theta(t_1,t_2)}$ is holomorphic and no zero points  in the   complex space, then the local attenuation $a(t_1,t_2)=\frac{1}{2}\ln(r^2+m^2)$ and the local phase
$\theta(x_1,x_2)=\arctan \left( \frac{m}{r}\right)$ are satisfied  the Cauchy-Riemann equations as following,
 we have
\begin{eqnarray*}\label{eq3}
\frac{\partial a}{\partial t_2}+\frac{\partial \theta}{\partial t_1}=0,
\end{eqnarray*}
\begin{eqnarray*}\label{eq4}
\frac{\partial a}{\partial t_1}-\frac{\partial \theta}{\partial t_2}=0.
\end{eqnarray*}
%From the above equations, we notice that:
%\begin{itemize}
%\item
%The partial derivative of the local attenuation  ${\partial a\over \partial t_2}$ in variable $t_2$
%equals the minus of the instantaneous frequency $\frac{\partial \theta}{\partial t_1}$ .
%\item the extreme of the local attenuation is  obtained by the zero points of the scale derivative of the local
%phase ${\partial \theta \over \partial t_2}$.\end{itemize}
%%===============================================================================
%
%Building on the ideas of 1D, Felsberg et al. \cite{felsberg2004monogenic}  proposed a edge detection approach based on  the phase of  {\it intrinsically 1D monogenic signals}, it is called
%the  \textbf{differential phase congruency (DPC)} method.
%%====================================================================
% While in higher dimension, the above  equations do  not hold in general. %===============================================================================
%In \cite{yang2017edge}, authors considered the monogenic signals in  higher dimension case, proposed a modified edge detection method which is denoted  \textbf{ modified differential phase congruency (MDPC)}.

In fact, if $g(s_1, s_2)=r(s_1, s_2)+\underline{m}(s_1, s_2)=e^{a(s_1,s_2)+\underline{p}(s_1, s_2)}$ is  holomorphic respective to $s_1,s_2$
in the upper half space $\mathbb{CS}$, in
general, $a(s_1, s_2)+\underline{p}(s_1, s_2)$ is not holomorphic.
%--------------------------------------------
\begin{Exa} Let $E(s_1,s_2)={s_{1}^{\ast}s_{2}^{\ast} \over |s_{1}|^{2} |s_{2}|^{2}}$
be the Cauchy kernel in $\mathbb{H}\setminus \{0\}$, which is holomorphic in $\mathbb{H}\setminus \{0\}$. Then, by straightforward computations, we have
\begin{eqnarray*}
&&a(s_1,s_2)+\underline{p}(z_1,z_2)\\
&=&\ln(\sqrt{t^2_{1}+y^2_{1}})+ \ln(\sqrt{t^2_{2}+y^2_{2}})+
\frac{-\i t_2y_1 -\j t_1 y_2 +\k y_1 y_2}{|-\i t_2y_1 -\j t_1 y_2 +\k y_1 y_2|}\\
&&\arctan \left(\frac{|-\i t_2y_1 -\j t_1 y_2 +\k y_1 y_2|}{t_1t_2}\right).
\end{eqnarray*}
 Applying  the generalized Cauchy-Riemann operators ${\partial \over \partial \overline{s_1}}$, ${\partial \over \partial \overline{s_2}}$  on it left side and right side, respectively,   we obtain %yields
\begin{eqnarray*}
 {\partial \over \partial \overline{s_1}} (a(s_1,s_2)+\underline{p}(s_1,s_2) )\neq 0. \quad (m(s_1,s_2)+\underline{p}(s_1,s_2) ) {\partial \over \partial \overline{s_2}} \neq 0.
\end{eqnarray*}
Therefore, $(a(s_1,s_2)+\underline{p}(s_1,s_2) )$ is not quaternion  holomorphic respective to $s_1,s_2.$
\end{Exa}
\begin{Prob}\label{P1} What is  relations of the local features of quaternion Hardy function in high dimensions ?\end{Prob}

%====================================================================
 Corollary  \ref{coro11} gives the answers for   Problem \ref{P1}. Theorem \ref{th1} shows the more detail relations
between   four components of $f$ and its local attenuation $a$ in higher
dimensional spaces. Before we give our main results, we first need some lemmas.
%----------------------------------------------------------------
\begin{Rem}
For convenience,
let $$  \underline{p}:=\underline{p}(s_1, s_2), \underline{m}:=\underline{m}(s_1, s_2), \theta:=\theta(s_1, s_2).
 a:=a(s_1, s_2), r:=r(s_1, s_2).$$

\end{Rem}
%-------------------------------------
\begin{Lem}\label{newth2}
Set $ \underline{p}(s_1, s_2)=\frac{\underline{m}(s_1, s_2)}{|\underline{m}(s_1, s_2)|}\theta(s_1, s_2),
$ which is defined in Eq. (\ref{lpv}),
 Then we obtain
\begin{eqnarray}\label{addeqx}
\frac{\partial \underline{p}}{\partial
t_{i}}=\left(\theta-\sin\theta\cos\theta\right)\frac{\partial
\frac{\underline{m}}{|\underline{m}|}}{\partial t_{i}}+
\frac{r\frac{\partial \underline{m}}{\partial
t_{i}}-\underline{m}\frac{\partial r}{\partial t_{i}}
}{r^2+|\underline{m}|^2}.
\end{eqnarray}
\begin{eqnarray}\label{addeqy}
\frac{\partial \underline{p}}{\partial
y_{i}}=\left(\theta-\sin\theta\cos\theta\right)\frac{\partial
\frac{\underline{m}}{|\underline{m}|}}{\partial y_{i}}+
\frac{r\frac{\partial \underline{m}}{\partial
y_{i}}-\underline{m}\frac{\partial r}{\partial y_{i}}
}{r^2+|\underline{m}|^2}.
\end{eqnarray}
\end{Lem}

%============================================================
\begin{itemize}
\item {\bf Proof: } By  Eq. (\ref{lpv}), we obtain
\begin{eqnarray}\label{eq99}
\frac{\partial \underline{p}}{\partial t_{i}}&=&\frac{\partial}
{\partial t_{i}}(\frac{\uv}{|\uv|}\theta)=\frac{\partial
\frac{\uv}{|\uv|}} {\partial
 t_{i}}\theta+\frac{\uv}{|\uv|}\frac{\partial} {\partial t_{i}}\theta.
\end{eqnarray}
By straightforward computation, we have

\begin{eqnarray}\label{eq102}
\frac{\uv}{|\uv|}\frac{\partial} {\partial
t_{i}}(\arctan \left(\frac{|\uv|}{r}\right))=\frac{r\frac{\partial \uv}{\partial
t_{i}}-\uv\frac{\partial r}{\partial
t_{i}}}{r^2+|\uv|^2}-\frac{r|\uv|}{r^2+|\uv|^2}\frac{\partial
\frac{\uv}{|\uv|}}{\partial t_{i}}.
\end{eqnarray}
Combining  Eq. (\ref{eq99}) and  Eq. (\ref{eq102}), we obtain Eq. (\ref{addeqx}).
Similarly, we have Eq. (\ref{addeqy}).
\end{itemize}

%-------------------------------------------------------------
\begin{Lem}\label{lem3}
Let
$f(s_1,s_2)=r(s_1,s_2)+\underline{m}(s_1,s_2)=e^{a(s_1,s_2)+\underline{p}(s_1,s_2)} \in \mathbb{QS}(\mathbb{CS}, \mathbb{H})$, where $a(s_1,s_2)$ and $\ur(s_1,s_2)$ are  defined by (\ref{atten}) and (\ref{lpv}), respectively. If $f$ has no zeros in the half space $\mathbb{CS}$. Then we have
\begin{eqnarray}\label{eglem1}
\left(\frac{\partial}{\partial t_{1}}e^{\underline{p}}\right)e^{-\underline{p}}=\frac{u\frac{\partial \underline{m}}{\partial
t_{1}}-\underline{m}\frac{\partial r}{\partial t_{1}}
}{r^2+|\underline{m}|^2}-\sin^2\theta\frac{\partial \frac{\uv}{|\uv|}}{\partial
t_1}\frac{\uv}{|\uv|}.
\end{eqnarray}
\begin{eqnarray}\label{eglem2}
\left(\frac{\partial}{\partial y_{1}}e^{\underline{p}}\right)e^{-\underline{p}}=\frac{r\frac{\partial \underline{m}}{\partial
y_{1}}-\underline{m}\frac{\partial r}{\partial y_{1}}
}{r^2+|\underline{m}|^2}-\sin^2\theta\frac{\partial \frac{\uv}{|\uv|}}{\partial
y_1}\frac{\uv}{|\uv|}.
\end{eqnarray}
\begin{eqnarray}\label{eglem3}
e^{-\underline{p}}\left(\frac{\partial}{\partial t_{2}}e^{\underline{p}}\right)=
\frac{r\frac{\partial \underline{m}}{\partial
t_{i}}-\underline{m}\frac{\partial r}{\partial t_{i}}
}{r^2+|\underline{m}|^2}-\frac{\uv}{|\uv|}\sin^2\theta\frac{\partial \frac{\uv}{|\uv|}}{\partial
t_2}.
\end{eqnarray}
\begin{eqnarray}\label{eglem4}
e^{-\underline{p}}\left(\frac{\partial}{\partial y_{2}}e^{\underline{p}}\right)=
\frac{r\frac{\partial \underline{m}}{\partial
y_{2}}-\underline{m}\frac{\partial r}{\partial y_{2}}
}{r^2+|\underline{m}|^2}-\frac{\uv}{|\uv|}\sin^2\theta\frac{\partial \frac{\uv}{|\uv|}}{\partial
y_2}.
\end{eqnarray}
\end{Lem}

%----------------------------------
\begin{itemize}\item {\bf Proof: }

 Using the generalized Euler formula
$e^{\underline{p}}=e^{\frac{\underline{m}}{|\underline{m}|}\theta}=\cos\theta+\frac{\uv}{|\uv|} \sin
\theta$, we have
\begin{eqnarray}\label{qq}
&& \left(\frac{\partial}{\partial t_{1}}e^{\ur}\right)  e^{-\ur}\nonumber\\
&=&\frac{\partial}{\partial t_1} \left(\cos \theta+\frac{\uv}{|\uv|}\sin\theta\right) \left(\cos\theta-\frac{\uv}{|\uv|}\sin\theta\right) \nonumber\\
&=&\left(-\sin\theta \frac{\partial \theta}{\partial
t_1}+\frac{\partial \frac{\uv}{|\uv|}}{\partial
t_1}\sin\theta+\frac{\uv}{|\uv|}\cos\theta \frac{\partial
\theta}{\partial t_1}\right)
\left(\cos\theta-\frac{\uv}{|\uv|}\sin\theta\right)\nonumber\\
&=&\frac{\uv}{|\uv|} \frac{\partial\theta}{\partial
t_1}+\sin\theta\cos\theta \frac{\partial \frac{\uv}{|\uv|}}{\partial
t_1} -\sin^2\theta\frac{\partial \frac{\uv}{|\uv|}}{\partial
t_1}\frac{\uv}{|\uv|}.
\end{eqnarray}

From (\ref{qq}), we know that
$(\frac{\partial}{\partial t_1}e^{\underline{p}})e^{-\underline{p}}$ is decided by $\frac{\uv}{|\uv|}
\frac{\partial\theta}{\partial t_1}+\sin\theta\cos\theta
\frac{\partial \frac{\uv}{|\uv|}}{\partial t_1}-\sin^2\theta\frac{\partial \frac{\uv}{|\uv|}}{\partial
t_1}\frac{\uv}{|\uv|}$. Since $\underline{p}=\frac{\uv}{|\uv|}\theta$, we have

\begin{equation}\label{qq1}
\frac{\partial \underline{p}}{\partial t_1}=\frac{\partial\theta}{\partial
t_1}\frac{\uv}{|\uv|}+\theta\frac{\partial \frac{\uv}{|\uv|}}{\partial
t_1}.
\end{equation}

 From  Eq.(\ref{addeqx}),  therefore, we obtain Eq.(\ref{eglem1}). Similarly,
we obtain (\ref{eglem2}),(\ref{eglem3}),(\ref{eglem4}).
\end{itemize}

%=====================================================================
%====================================================

The next theorem contains  one of main results.

\begin{Th}\label{th1}
Let $f(s_1,s_2)=r(s_1,s_2)+\underline{m}(s_1,s_2)=e^{a(s_1,s_2)+\underline{p}(s_1,s_2)} \in \mathbb{QS}(\mathbb{CS}, \mathbb{H})$, where $a(s_1,s_2)$ and $\ur(s_1,s_2)$ are defined by (\ref{atten}) and (\ref{lpv}), respectively. If $f$ has no zeros  in the upper half space $\mathbb{CS}$. Then we have
\begin{eqnarray}\label{Teq7}
\frac{\partial a}{\partial t_{1}}-\bigg\{\frac{u\frac{\partial m_{1}}{\partial
y_{1}}-m_{1}\frac{\partial r}{\partial y_{1}}
}{r^2+|\underline{m}|^2}-\sin^2\theta\frac{\partial \frac{m_{2}}{|\uv|}}{\partial
y_1}\frac{m_3}{|\uv|}+\sin^2\theta\frac{\partial \frac{m_{3}}{|\uv|}}{\partial
y_1}\frac{m_2}{|\uv|} \bigg\}=0.
\end{eqnarray}
\begin{eqnarray}\label{Teq8}
 \frac{\partial a}{\partial y_{1}}+\bigg\{\frac{r\frac{\partial m_{1}}{\partial
t_{1}}-m_{1}\frac{\partial r}{\partial t_{1}}
}{r^2(s_{1}, s_{2})+|\underline{m}(s_{1}, s_{2})|^2}-\sin^2\theta\frac{\partial \frac{m_{2}}{|\uv|}}{\partial
t_1}\frac{m_3}{|\uv|}+\sin^2\theta\frac{\partial \frac{m_{3}}{|\uv|}}{\partial
t_1}\frac{m_2}{|\uv|} \bigg\} =0.
\end{eqnarray}
\begin{eqnarray*}\label{Teq9}
{\rm
Vec(\j)}[\frac{\partial (e^{\underline{p}})e^{-\underline{p}}}{\partial t_{1}}+
\i \frac{\partial (e^{\underline{p}})e^{-\underline{p}}}{\partial y_{1}}] =0.
\end{eqnarray*}
\begin{eqnarray*}\label{Teq10}
{\rm
Vec(\k)}[\frac{\partial (e^{\underline{p}})e^{-\underline{p}}}{\partial t_{1}}+
\i \frac{\partial( e^{\underline{p}})e^{-\underline{p}}}{\partial y_{1}}] =0.
\end{eqnarray*}
\begin{eqnarray}\label{Teq11}
\frac{\partial a}{\partial t_{2}}-\bigg\{\frac{r\frac{\partial m_{2}}{\partial
y_{2}}-m_{2}\frac{\partial r(s_{1}, s_{2})}{\partial y_{2}}
}{r^2+|\underline{m}|^2}-\sin^2\theta\frac{\partial \frac{m_{1}}{|\uv|}}{\partial
y_2}\frac{m_3}{|\uv|}+\sin^2\theta\frac{\partial \frac{m_{3}}{|\uv|}}{\partial
y_2}\frac{m_1}{|\uv|} \bigg\}=0.
\end{eqnarray}
\begin{eqnarray}\label{Teq12}
\frac{\partial a}{\partial y_{2}}+\bigg\{\frac{r\frac{\partial m_{2}}{\partial
t_{2}}-m_{2}\frac{\partial r}{\partial t_{2}}
}{r^2+|\underline{m}|^2}-\sin^2\theta\frac{\partial \frac{m_{1}}{|\uv|}}{\partial
t_2}\frac{m_3}{|\uv|}+\sin^2\theta\frac{\partial \frac{m_{3}}{|\uv|}}{\partial
t_2}\frac{m_1}{|\uv|} \bigg\}=0=0,
\end{eqnarray}

\begin{eqnarray}\label{Teq13}
{\rm
Vec(\i)}[\frac{e^{-\underline{p}}\partial (e^{\underline{p}})}{\partial t_{2}}+
 \frac{e^{-\underline{p}}\partial (e^{\underline{p}})}{\partial y_{2}}\j] =0.
\end{eqnarray}

\begin{eqnarray}\label{Teq14}
{\rm
Vec(\k)}[\frac{e^{-\underline{p}}\partial (e^{\underline{p}})}{\partial t_{2}}+
 \frac{e^{-\underline{p}}\partial (e^{\underline{p}})}{\partial y_{2}}\j] =0.
\end{eqnarray}
\end{Th}

%======================================
\begin{itemize}\item
\noindent {\bf Proof of Theorem \ref{th1}: } Since $f(s_1,
s_2)=e^{m+\underline{p}} \in\mathbb{QS}(\mathbb{C}_{\i \j}^{+}))$, we have
$$\left(\frac{\partial}{\partial t_1}+\i\frac{\partial}{\partial y_1}\right) e^{a+\underline{p}}=0.$$ By straightforward computation, we have
$$e^{a}\frac{\partial a}{\partial t_1}e^{\underline{p}}+e^{a}\frac{\partial e^{\ur}}{\partial t_1}
+e^{a}[\i\frac{\partial a}{\partial y_1}]e^{\ur}+e^{a}(\i\frac{\partial e^{\ur}}{\partial y_1})=0.$$ That is
\begin{eqnarray}\label{eq13}
\frac{\partial a}{\partial t_1}+\frac{\partial e^{\ur}}{\partial t_1}e^{-\ur} +\i\frac{\partial a}{\partial y_1}+(\i\frac{\partial e^{\ur}}{\partial y_1})e^{-\ur}=0.
\end{eqnarray}
Therefore, the scale part of (\ref{eq13}) is zero, and
$ \partial\frac{ \left(\frac{\underline{m}}{|m|}\right)^2 }{ \partial t_1}=0,$
 we have
\begin{eqnarray}
&&{\rm Sc}\left[\frac{\partial a}{\partial t_1}+\frac{\partial e^{\ur}}{\partial t_1}e^{-\ur} +\i\frac{\partial m}{\partial y_1}+(\i\frac{\partial e^{\ur}}{\partial y_1})e^{-\ur}\right]\nonumber\\
&=&\frac{\partial a}{\partial t_1}+{\rm Sc}[(\i\frac{\partial e^{\ur}}{\partial y_1})e^{-\ur}]=0\label{hxx1}
\end{eqnarray}
Moreover, from (\ref{eglem1}),
\begin{eqnarray*}
{\rm
Sc}\bigg[\bigg(\i \frac{\partial (e^{\underline{p}})e^{-\underline{p}}}{\partial y_{1}}\bigg)\bigg]
=-\bigg\{\frac{r\frac{\partial m_{1}}{\partial
y_{1}}-m_{1}\frac{\partial r}{\partial y_{1}}
}{r^2+|\underline{m}|^2}-\sin^2\theta\frac{\partial \frac{m_{2}}{|\uv|}}{\partial
y_1}\frac{m_3}{|\uv|}+\sin^2\theta\frac{\partial \frac{m_{3}}{|\uv|}}{\partial
y_1}\frac{m_2}{|\uv|} \bigg\}.
\end{eqnarray*}
we get the desired result (\ref{Teq7}).

%\begin{eqnarray}\label{hxx2}
%{\rm
%Re}\bigg[\bigg( e^{-\underline{p}}\frac{\partial (e^{\underline{p}})}{\partial y_{2}}\j\bigg)\bigg]=
%-\bigg\{\frac{u(z_{1}, z_{2})\frac{\partial v_{2}(z_{1}, z_{2})}{\partial
%y_{2}}-v_{2}(z_{1}, z_{2})\frac{\partial u(z_{1}, z_{2})}{\partial y_{2}}
%}{u^2(z_{1}, z_{2})+|\underline{v}(z_{1}, z_{2})|^2}-\sin^2\theta\frac{\partial \frac{v_{1}}{|\uv|}}{\partial
%y_2}\frac{v_3}{|\uv|}+\sin^2\theta\frac{\partial \frac{v_{3}}{|\uv|}}{\partial
%y_2}\frac{v_1}{|\uv|} \bigg\}.
%\end{eqnarray}
The three vector parts of  Eq. (\ref{eq13}) are  all  zero. Using Lemma \ref{lem3} we find that
\begin{eqnarray}\label{eq15}
&&{\rm Vec(\i)}\left[\frac{\partial a}{\partial t_1}+\frac{\partial e^{\ur}}{\partial t_1}e^{-\ur} +\i\frac{\partial a}{\partial y_1}+(\i\frac{\partial e^{\ur}}{\partial y_1})e^{-\ur}\right]\nonumber\\
&=&{\rm Vec(\i)}\left[\frac{\partial e^{\ur}}{\partial t_1}e^{-\ur} +\i\frac{\partial a}{\partial y_1}\right]\nonumber\\
&=& \frac{\partial a}{\partial y_1}+{\rm Sc}\left[ (-\i)\frac{\partial e^{\ur}}{\partial t_1}e^{-\ur}\right]
=0\label{re2}.
\end{eqnarray}
from (\ref{eglem2}), we  obtain Eq.(\ref{Teq8}).

\begin{eqnarray}\label{eq15}
&&{\rm Vec(\j)}\left[\frac{\partial a}{\partial t_1}+\frac{\partial e^{\ur}}{\partial t_1}e^{-\ur} +\i\frac{\partial a}{\partial y_1}+(\i\frac{\partial e^{\ur}}{\partial y_1})e^{-\ur}\right]\nonumber\\
&=&{\rm Vec(\j)}\left[\frac{\partial e^{\ur}}{\partial t_1}e^{-\ur} +(\i\frac{\partial e^{\ur}}{\partial y_1})e^{-\ur}\right]
=0\nonumber.
\end{eqnarray}
\begin{eqnarray}\label{eq15}
&&{\rm Vec(\k)}\left[\frac{\partial a}{\partial t_1}+\frac{\partial e^{\ur}}{\partial t_1}e^{-\ur} +\i\frac{\partial a}{\partial y_1}+(\i\frac{\partial e^{\ur}}{\partial y_1})e^{-\ur}\right]\nonumber\\
&=&{\rm Vec(\k)}\left[\frac{\partial e^{\ur}}{\partial t_1}e^{-\ur} +(\i\frac{\partial e^{\ur}}{\partial y_1})e^{-\ur}\right]
=0\nonumber.
\end{eqnarray}
Since $$ e^{a+\underline{p}}\left(\frac{\partial}{\partial t_2}+\j\frac{\partial}{\partial y_2}\right)=0.$$
Using the similar steps as in above, we have
\begin{eqnarray}\label{re3}
\frac{\partial a}{\partial t_2}+{\rm Sc}\left[e^{-\ur}(\frac{\partial e^{\ur}}{\partial y_2})\j\right]=0
\end{eqnarray}
\begin{eqnarray}\label{re4}
\frac{\partial a}{\partial y_2}-{\rm Sc}\left[e^{-\ur}(\frac{\partial e^{\ur}}{\partial t_2})\j\right]=0
\end{eqnarray}
\begin{eqnarray*}
{\rm Vec(\i)}\left[(e^{-\ur}\frac{\partial e^{\ur}}{\partial t_2})+e^{-\ur}(\frac{\partial e^{\ur}}{\partial y_2})\j\right]=0
\end{eqnarray*}
\begin{eqnarray*}
{\rm Vec(\k)}\left[e^{-\ur}(\frac{\partial e^{\ur}}{\partial t_2})+e^{-\ur}(\frac{\partial e^{\ur}}{\partial y_2})\j\right]=0
\end{eqnarray*}

Using  Lemma \ref{lem3} and  Eqs.(\ref{eglem3}), (\ref{eglem4}), we obtain Eqs.(\ref{Teq11}),(\ref{Teq12}),(\ref{Teq13}),(\ref{Teq14}).This completes the proof.
\end{itemize}
%========================

From Theorem \ref{th1}, combing the Eqs.(\ref{hxx1}), (\ref{re2}), (\ref{re3}), (\ref{re4}), (\ref{qq}), (\ref{qq1}),
 we have the following  corollary, which show the relations of the local
features of quaternion
Hardy function.
%===============================
\begin{Cor}\label{coro11}
Let $f(s_1,s_2)=r(s_1,s_2)+\underline{m}(s_1,s_2)=e^{a(s_1,s_2)+\underline{p}(s_1,s_2)} \in \mathbb{QS}(\mathbb{C}, \mathbb{H})$,  If $f$ has no zeros  in the upper half space $\mathbb{CS}$. Then we have
\begin{eqnarray}\label{coroe1}
\frac{\partial a}{\partial t_1}+{\rm Vec(\i)}\left[\frac{\partial \underline{p}}{\partial y_1}-\theta\frac{\partial \frac{\uv}{|\uv|}}{\partial
y_1}+\sin\theta\cos\theta \frac{\partial \frac{\uv}{|\uv|}}{\partial
y_1} -\sin^2\theta\frac{\partial \frac{\uv}{|\uv|}}{\partial
y_1}\frac{\uv}{|\uv|}\right]=0
\end{eqnarray}
\begin{eqnarray}\label{coroe2}
\frac{\partial a}{\partial t_2}+{\rm Vec(\j)}\left[\frac{\partial \underline{p}}{\partial y_2}-\theta\frac{\partial \frac{\uv}{|\uv|}}{\partial
y_2}+\sin\theta\cos\theta \frac{\partial \frac{\uv}{|\uv|}}{\partial
y_2} -\frac{\uv}{|\uv|}\sin^2\theta\frac{\partial \frac{\uv}{|\uv|}}{\partial
y_2}\right]=0
\end{eqnarray}
\begin{eqnarray}\label{coroe3}
\frac{\partial a}{\partial y_1}-{\rm Vec(\i)}\left[\frac{\partial \underline{p}}{\partial t_1}-\theta\frac{\partial \frac{\uv}{|\uv|}}{\partial
t_1}+\sin\theta\cos\theta \frac{\partial \frac{\uv}{|\uv|}}{\partial
t_1} -\sin^2\theta\frac{\partial \frac{\uv}{|\uv|}}{\partial
t_1}\frac{\uv}{|\uv|}\right]=0
\end{eqnarray}

\begin{eqnarray}\label{coroe4}
\frac{\partial a}{\partial y_2}-{\rm Vec(\j)}\left[\frac{\partial \underline{p}}{\partial t_2}-\theta\frac{\partial \frac{\uv}{|\uv|}}{\partial
t_2}+\sin\theta\cos\theta \frac{\partial \frac{\uv}{|\uv|}}{\partial
t_2} -\frac{\uv}{|\uv|}\sin^2\theta\frac{\partial \frac{\uv}{|\uv|}}{\partial
t_2}\right]=0
\end{eqnarray}

\end{Cor}
From  this corollary,  we also notice that
\begin{itemize}
\item Eqs. \ref{coroe1} and \ref{coroe2}  show that the partial  derivatives of the local attenuation  $\frac{\partial a}{\partial t_1}$ and $\frac{\partial a}{\partial t_2}$ do  not equal
the  the minus of the scale derivatives of the local
phase-vector ${\partial\underline{ p }\over \partial y_1} $ and ${\partial\underline{ p} \over \partial y_2} $.
\item Eqs. \ref{coroe3} and \ref{coroe4} show that  the partial derivative of the local
phase-vector ${\partial \underline{ p }\over \partial t_1}$ and ${\partial \underline{ p }\over \partial t_2}$ can not be  given by the scale partial  derivatives of the local attenuation $\frac{\partial a}{\partial y_1}$ and $\frac{\partial a}{\partial y_2}.$
\end{itemize}

%%%%%%%%%%%%%%%%%%%%%%%%%%%%%%%%%%%%%%%%%
\subsection{Concrete  Approaches}\label{hxx4}

Let's first introduce the Quaternion Differential Local Attenuation (QDLA) approach,
which is based on the partial differentiation  of the local attenuation in the $ t_1$ and $t_2$ variables.
\begin{Method}[QDLA]
For $f=r+\i m_1 +\j m_2 +\k m_3 \in\mathbb{QS}(\mathbb{CS},\mathbb{ H})$ has no zeros  in the half space $\mathbb{CS}$,  the Quaternion Differential Local Attenuation (QDLA) approach has the formula
\begin{eqnarray*}\label{LAM}
\frac{\partial a}{\partial t_1}&=&\frac{u\frac{\partial r}{\partial t_1}+|\underline{m}|\frac{\partial |\underline{m}|}{\partial t_1}}{r^2+|\underline{m}|^2}.\\
\frac{\partial a}{\partial t_2}&=&\frac{r\frac{\partial r}{\partial t_2}+|\underline{m}|\frac{\partial |\underline{m}|}{\partial t_2}}{r^2+|\underline{m}|^2}.\\
\end{eqnarray*}
\end{Method}

From Eqs. (\ref{Teq7}) and (\ref{Teq11}),  the local maxima of the local amplitude is equivalent to
Eqs. (\ref{MDLA1}) and (\ref{MDLA2}) for QHF, then we propose the next approach, named  Modified  Quaternion Differential Local Attenuation (MQDLA), which based on the local phase.
%%%%%%%%%%%%%%%%%%%%%%%%%%%%%%%%%%%%%%%%%%%%%%%%%%%%%%%%%%%%%%%%%%%%%%%%%%%%%
\begin{Method}[MQDLA]
For $f=r+\i m_1 +\j m_2 +\k m_3 \in\mathbb{QS}(\mathbb{CS},\mathbb{ H})$ has no zeros  in the half space $\mathbb{CS}$, the  MQDLA approach has the formula

\begin{eqnarray}\label{MDLA1}
p1=\frac{r\frac{\partial m_{1}}{\partial
y_{1}}-m_{1}\frac{\partial r}{\partial y_{1}}
}{r^2+|\underline{m}|^2}-\sin^2\theta\frac{\partial \frac{m_{2}}{|\uv|}}{\partial
y_1}\frac{m_3}{|\uv|}+\sin^2\theta\frac{\partial \frac{m_{3}}{|\uv|}}{\partial
y_1}\frac{m_2}{|\uv|}.
\end{eqnarray}
\begin{eqnarray}\label{MDLA2}
p2=\frac{r\frac{\partial m_{2}}{\partial
y_{2}}-m_{2}\frac{\partial r}{\partial y_{2}}
}{r^2+|\underline{m}(s_{1}, s_{2})|^2}-\sin^2\theta\frac{\partial \frac{m_{1}}{|\uv|}}{\partial
y_2}\frac{m_3}{|\uv|}+\sin^2\theta\frac{\partial \frac{m_{3}}{|\uv|}}{\partial
y_2}\frac{m_1}{|\uv|}.
\end{eqnarray}
\end{Method}

%%%%%%%%%%%%%%%%%%%%%%%%%%%%%%%%%%%%%%%%%%%%%%%%%%%%%%%%%%%%%%%%%%%%%%%%%%%%%%%%%%%%%%%%%%%%%
Then local attenuation has four  variables, let us derivative the local attenuation with respect to the scales $ y_1, y_2 $ to get the
  Scale Derivative Local Amplitude (SDLA) approach, which bases on the local amplitude.

%%%%%%%%%%%%%%%%%%%%%%%%%%%%%%%%%%%%%%%%%%%%%%%%%%%%%%%%%%%%%%%%%%%%%%%%%%%%%%%%%%%%%%%%
\begin{Method}[SDLA]
For $f=r+\i m_1 +\j m_2 +\k m_3 \in\mathbb{QS}(\mathbb{CS},\mathbb{ H})$ has no zeros  in the half space $\mathbb{CS}$,  the SDLA approach has the formula
\begin{eqnarray*}\label{LAM}
\frac{\partial a}{\partial y_1}&=&\frac{u\frac{\partial r}{\partial y_1}+|\underline{m}|\frac{\partial |\underline{m}|}{\partial y_1}}{r^2+|\underline{m}|^2}.\\
\frac{\partial a}{\partial y_2}&=&\frac{r\frac{\partial r}{\partial y_2}+|\underline{m}|\frac{\partial |\underline{m}|}{\partial y_2}}{r^2+|\underline{m}|^2}.\\
\end{eqnarray*}
\end{Method}
%%%%%%%%%%%%%%%%%%%%%%%%%%%%%%%%%%%%%%%%%%%%%%%%%%%%%%%%%%%%%%%%%%%%%%%%%%%%%%%%%%%%%%%%%%%%%
 From Eqs. (\ref{Teq8}) and (\ref{Teq12}),  the scale partial  derivatives of the local attenuation$\frac{\partial a}{\partial y_1}$ and $\frac{\partial a}{\partial y_2}$ are equivalent to
Eqs. (\ref{mqlay1}) and (\ref{mqlay2}) for QHF, respectively,  then we propose the next approach, named Modified Scale Derivative Local Amplitude (MDSLA) approach, which bases on the local phase.
%%%%%%%%%%%%%%%%%%%%%%%%%%%%%%%%%%%%%%%%%%%%%%%%%%%%%%%%%%%%%%%%%%%%%%%%%%%%%%%%%%%%%%%%%%
\begin{Method}[MSDLA]
For $f=r+\i m_1 +\j m_2 +\k m_3 \in\mathbb{QS}(\mathbb{CS},\mathbb{ H})$ has no zeros  in the half space $\mathbb{CS}$, , the MSDLA approach has the formula

\begin{eqnarray}\label{mqlay1}
M1:=\frac{-r\frac{\partial m_{1}}{\partial
t_{1}}+m_{1}\frac{\partial r}{\partial t_{2}}
}{r^2+|\underline{m}|^2}+\sin^2\theta\frac{\partial \frac{m_{2}}{|\uv|}}{\partial
t_1}\frac{m_3}{|\uv|}-\sin^2\theta\frac{\partial \frac{m_{3}}{|\uv|}}{\partial
t_1}\frac{m_2}{|\uv|}
\end{eqnarray}
\begin{eqnarray}\label{mqlay2}
M2:=\frac{-r\frac{\partial m_{2}}{\partial
t_{2}}+m_{2}\frac{\partial r}{\partial t_{2}}
}{r^2+|\underline{m}|^2}+\sin^2\theta\frac{\partial \frac{m_{1}}{|\uv|}}{\partial
t_2}\frac{m_3}{|\uv|}-\sin^2\theta\frac{\partial \frac{m_{3}}{|\uv|}}{\partial
t_2}\frac{m_1}{|\uv|}
\end{eqnarray}
\end{Method}

%%%%%%%%%%%%%%%%%%%%%%%%%%%%%%%%%%%%%%%%%%%%%%%%%%%%%%%%%%%%%%%%%%%%%%%%%%%%%%%%%%%%%%%%

\section{Experiments}\label{hxx5}
In this section,
 In order to have an idea of the performance, the new
approaches are not only compared to edge detection algorithms taken from the image processing
toolbox of Matlab: Canny and Sobel edge detectors, but also compared  ours to the phase congruency edge detectors, DPC \cite{felsberg2004monogenic} and MDPC \cite{yang2017edge}. The Comparative results of the proposed  approaches $($ QDLA, MQDLA, SDLA, MSDLA $)$ and the Canny , Sobel, DPC, MDPC approachs  are visualized in Fig. \ref{fig1}.

The approach of MDPC and DPC is used the Non-maximum suppress \cite{kovesi1999image} to get thinner edge boundary,the radius is chosen r=1.5,  for the MDPC,  the lower and upper thresholds  are  1.0 and 3.5, respectively. While for the DPC, the lower and upper thresholds  are 2.0 and 3.5, respectively.
And all parameters of Canny and Sobel are  automatically generated by the Matlab
algorithms in order to obtain a comparison of fully automatic approaches.
%For our proposed approaches, we have the following Algorithms, for simply, the color image is converted to the gray image.

\subsection{Algorithms }
\begin{figure}[!]
 \centering
 \includegraphics[height=12cm,width=10cm]{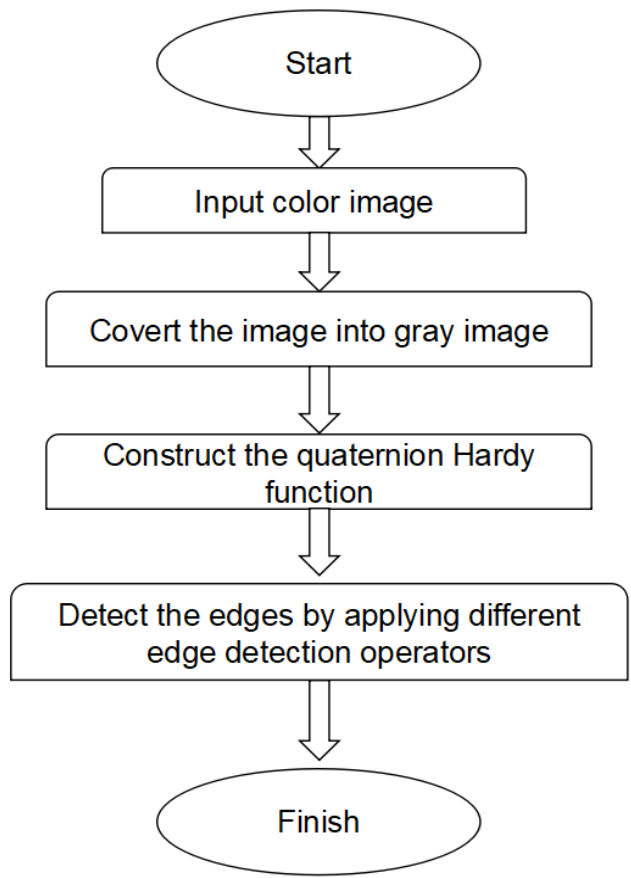}
  \caption{ Flow chart for image edge detection.}
  \label{fig00}
\end{figure}

%==========================
The flow of the image edge detection is shown in Figure \ref{fig00} .
%==========================
\begin{description}
  \item[Step 1.]
  Input image $r(t_1,t_2)$.
  \item[Step 2.]
   Poisson filtering \label{possion}: $r(t_1+\i y_1,t_2+\j y_2)=r*P_{y_{1}}(t_{1})P_{y_{2}}(t_{2})$ and and
   $ m_{1}(t_1+\i y_1,t_2+\j y_2)=r*Q_{y_{1}}(x_{1})P_{y_{2}}(t_{2}), $
   $m_{2}(t_1+\i y_1,t_2+\j y_2)=r*P_{y_{1}}(x_{1})Q_{y_{2}}(t_{2}),$
   $m_{3}(t_1+\i y_1,t_2+\j y_2)=r*Q_{y_{1}}(x_{1})Q_{y_{2}}(t_{2}), $
    for  fixed scales $ y_{1}=0.3, y_{1}=0.3 $.
    From this step, we obtain the quaternion Hardy function
     $$ r_{q}(s_1,s_2)=r(s_1,s_2)+\i m_{1}(s_1,s_2) + m_{2}(s_1,s_2)\j + \i m_{3}(s_1,s_2)\j. $$
  \item[Step 3.]
 Select  one of the algorithms to evaluate to obtain the gradient maps of these approach.:
 \par
 QDLA approach
 \begin{eqnarray*}\label{LAM}
\frac{\partial a}{\partial t_1}&=&\frac{r\frac{\partial r}{\partial t_1}+|\underline{m}|\frac{\partial |\underline{m}|}{\partial t_1}}{r^2+|\underline{m}|^2}.\\
\frac{\partial a}{\partial t_2}&=&\frac{r\frac{\partial r}{\partial t_2}+|\underline{m}|\frac{\partial |\underline{m}|}{\partial t_2}}{r^2+|\underline{m}|^2}.\\
\end{eqnarray*}
 MQDLA approach
 \begin{eqnarray*}\label{MDLA1}
p1=\frac{r\frac{\partial m_{1}}{\partial
y_{1}}-m_{1}\frac{\partial u}{\partial y_{1}}
}{r^2+|\underline{m}|^2}-\sin^2\theta\frac{\partial \frac{m_{2}}{|\uv|}}{\partial
y_1}\frac{m_3}{|\uv|}+\sin^2\theta\frac{\partial \frac{m_{3}}{|\uv|}}{\partial
y_1}\frac{m_2}{|\uv|}.
\end{eqnarray*}
\begin{eqnarray*}\label{MDLA2}
p2=\frac{r\frac{\partial m_{2}}{\partial
y_{2}}-m_{2}\frac{\partial r}{\partial y_{2}}
}{r^2+|\underline{m}|^2}-\sin^2\theta\frac{\partial \frac{m_{1}}{|\uv|}}{\partial
y_2}\frac{m_3}{|\uv|}+\sin^2\theta\frac{\partial \frac{m_{3}}{|\uv|}}{\partial
y_2}\frac{m_1}{|\uv|}.
\end{eqnarray*}
  SDLA approach
  \begin{eqnarray*}\label{LAM}
\frac{\partial a}{\partial y_1}&=&\frac{r\frac{\partial r}{\partial y_1}+|\underline{m}|\frac{\partial |\underline{m}|}{\partial y_1}}{r^2+|\underline{m}|^2}.\\
\frac{\partial a}{\partial y_2}&=&\frac{r\frac{\partial r}{\partial y_2}+|\underline{m}|\frac{\partial |\underline{m}|}{\partial y_2}}{r^2+|\underline{m}|^2}.\\
\end{eqnarray*}
 MSDLA approach
  \begin{eqnarray*}\label{mqlay1}
M1:=\frac{-r\frac{\partial m_{1}}{\partial
t_{1}}+m_{1}\frac{\partial r}{\partial t_{2}}
}{r^2+|\underline{m}|^2}+\sin^2\theta\frac{\partial \frac{m_{2}}{|\uv|}}{\partial
t_1}\frac{m_3}{|\uv|}-\sin^2\theta\frac{\partial \frac{m_{3}}{|\uv|}}{\partial
t_1}\frac{m_2}{|\uv|}
\end{eqnarray*}
\begin{eqnarray*}\label{mqlay2}
M2:=\frac{-r\frac{\partial m_{2}}{\partial
t_{2}}+m_{2}\frac{\partial r}{\partial t_{2}}
}{r^2+|\underline{m}|^2}+\sin^2\theta\frac{\partial \frac{m_{1}}{|\uv|}}{\partial
t_2}\frac{m_3}{|\uv|}-\sin^2\theta\frac{\partial \frac{m_{3}}{|\uv|}}{\partial
t_2}\frac{m_1}{|\uv|}
\end{eqnarray*}

  \item[Step 4.]
 For test images in  Fig.\ref{fig0} ,  the non-maximum suppress is applied to these gradient maps such that  the edges  become thinner,  the radius r=1.5 is chosen. For MSDLA detector, the lower and upper threshold values are 15 and 27, respectively.
For the other approaches, the lower and upper threshold values are 3.8 and 5.5, respectively.
\end{description}
%%%%%%%%%%%%%%%%%%%%%%%%%%%%%%%%%%%%%%%%%%%%%%%%%%%%%%%%%%%%%%%%%%%%%%%%%%%%%%%%%%%%%%%%%%%%%%%%%%%%%%%%%%
\subsection{Experiment results}

\subsubsection{Visual comparisons}
\begin{enumerate}
  \item There are four  original test  images, namely fish,  building, lane and liver (Figure \ref{fig0}), are applied for various edge detectors. The experiment results of various approaches  with the fixed scales $y_1=0.3$ and $y_2=0.3$ are shown in  Figure \ref{fig1} .
   Some conclusions  are reasonably  drawn from these results (Figure \ref{fig1}).
%==========================

\begin{figure}[!]
 \centering
 \includegraphics[height=4cm,width=14cm]{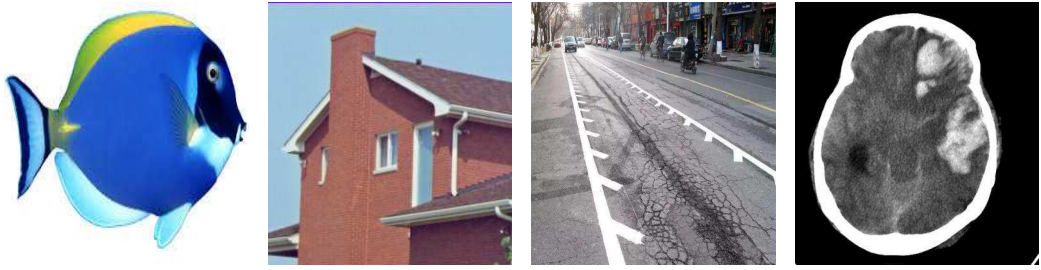}
  \caption{ The original test images : fish, building, lane and liver.}
  \label{fig0}
\end{figure}
%==========================
\begin{figure}[!h]
  \centering
 \includegraphics[height=23cm, width=14cm]{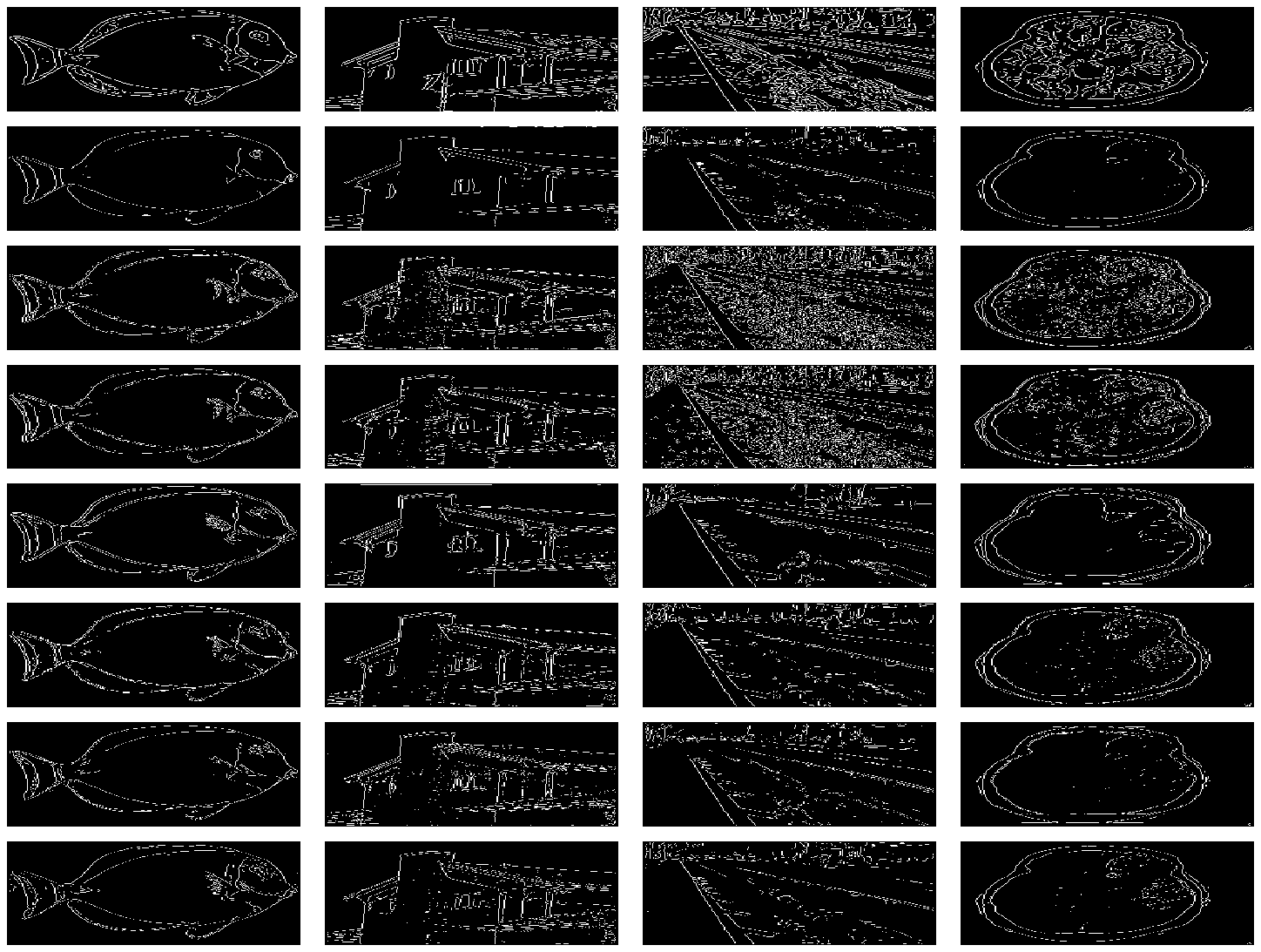}
  \caption{Comparative Results of Candy, Sobel, DPC, MDPC, QDLA,MQDLA, SDLA, MSDLA  approaches, from top to the bottom. }
  \label{fig1}
\end{figure}
%%%%%%%%%%%%%%%%%%%%%%%%%%

  \begin{itemize}

    \item  Firstly, Fig \ref{fig1} show us that the phase based approaches DPC,  MDPC and  our approaches, SDLA, MQDLA, MSDLA can achieve good performance in dealing with details. They can detect  a  pectoral fin of the fish,
         shadow region of the building and the white area of the liver.  The canny  and Sobel did not detect them,
         because Canny  regard them as noise and denoise  them. While the Sobel can not find them, because
         the color contrast between between the shadow region, pectoral fin and their surrounding area is  relatively small.
         \item Secondly, for the lane, our approaches and Sobel  can detect the lane lines very well, while DPC,MDPC and Candy  detect too much other information, hence can not figure out the lane lines.
        \item   Thirdly,  the MSDLA  has the best performance of all approaches visually.

      \end{itemize}

\begin{table}[h!]
\caption{Scales values $y_1$, $y_2$ and $s$ for (SDLA,MSDLA,QDLA,MQDLA) and (MDPC, DPC) approaches, respectively. }\label{tab1}
\centering
\begin{tabular}{|c|c|c|}
  \hline
  Noise & SDLA, MSDLA, QDLA, MQDLA      & MDPC, DPC \\

   \hline
  Possion noise  & $y_1=4.5,$ $y_2=4.5$    &  $s=3.5$ \\
   \hline
 Gaussian  noise &  $y_1=4.5,$ $y_2=4.5$    & $s=4.5$ \\
   \hline
   Salt and  pepper noise  & $y_1=5.8,$ $y_2=5.8$    &   $s=6$  \\
  \hline
  Speckle noise  & $y_1=4.5,$ $y_2=4.5$  &  $s=5.5$  \\
  \hline

 % \hline
%  Possion noise  & $y_1=4.5,$ $y_2=4.5$    &  $s=3.5$ \\
%   \hline
% Gaussian  noise
%with $0$ mean and $0.01$ variance &  $y_1=4.5,$ $y_2=4.5$    & $s=4.5$ \\
%   \hline
%   Salt and  pepper noise  with $0.05$ density & $y_1=5.8,$ $y_2=5.8$    &   $s=6$  \\
%  \hline
%  Speckle noise with  $0$ mean and $0.04$ variance  & $y_1=4.5,$ $y_2=4.5$  &  $s=5.5$  \\
%  \hline

\end{tabular}
 \end{table}
\item  Four  different kinds  of noise  are added to the test images, the scales values of the phased-based approaches  (MQDLA, MSDLA, DPC, MDPC) and  amplitude-based aproaches (QDLA, SDLA) are showed in Table \ref{tab1} shows. The SNR denotes the " signal to noise ratio ". Figs.\ref{fign3}, \ref{fign4} and \ref{fign5} show the comparative results. We can have the following conclusion.
\begin{itemize}
\item The MSDLA and SDLA  approaches give the better performance than other approaches  in detecting edge  from the visual comparison.
The MSDLA still detect clearly the internal edges of the images corrupted   Possion and  Gaussian noises, respectively.
The Our approaches  give the better performance in detecting the  noised lane lines  than others.
\end{itemize}

\end{enumerate}

%%%%%%%%%%%%%%%%%%%%%%%%%%%%%%%%%%%%%%%%%
%==========================
\begin{figure}[!h]
  \centering
 \includegraphics[height=13cm, width=14cm]{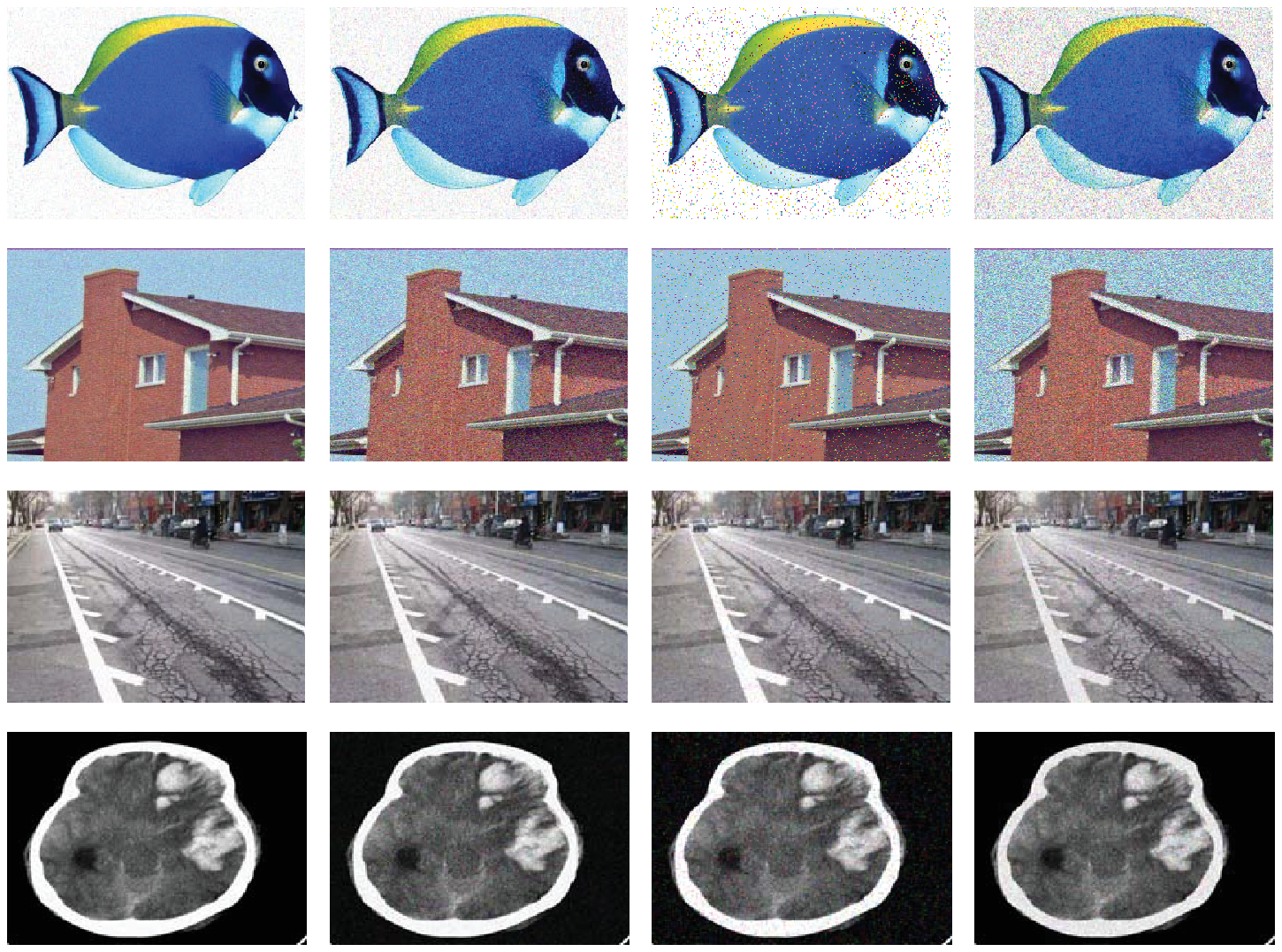}
  \caption{Adding from left to right by Possion, Gaussian, Salt and pepper, Speckle noises on  test images. }
  \label{fign2}
\end{figure}

%%%%%%%%%%%%%%%%%%%%%%%%%%%%%%%%%%%%%%%%%%%%%%%%%%%%%%%%%%

%%%%%%%%%%%%%%%%%%%%%%%%%%%%%%%%%%%%%%%%%%%%%%%%%%%%%%%%%%%%%%%%%%%
%==========================
\begin{figure}[!h]
  \centering
 \includegraphics[height=20cm, width=13cm]{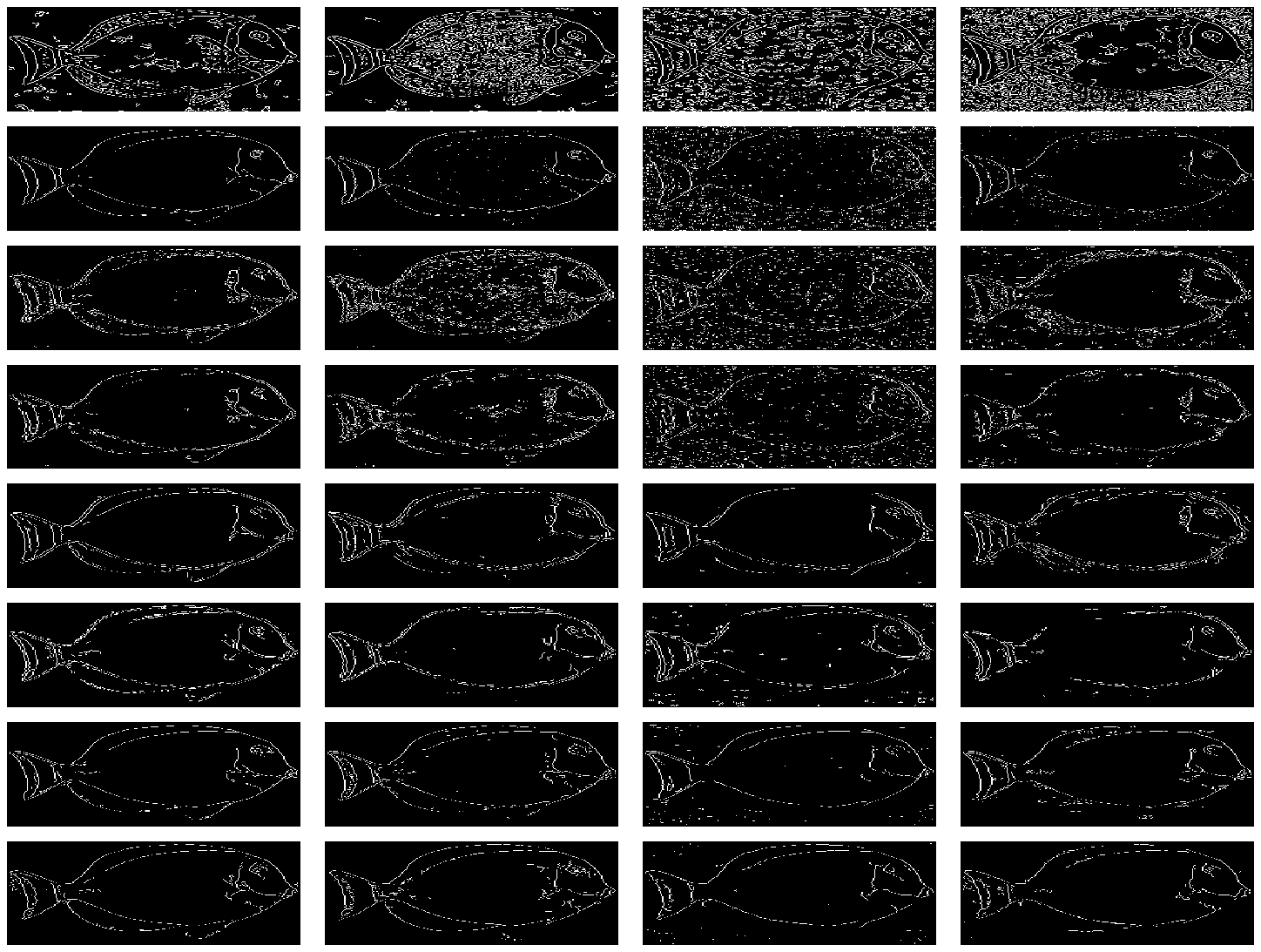}
  \caption{Comparative Results of Candy, Sobel, DPC, MDPC,  QDLA,MQDLA, SDLA, MSDLA  approaches on  fish image from top to the bottom. The first column is added by the  Possion, the second is added by the  Gaussian, the third column is added by the Salt and pepper, the last column is added by the  Speckle noises.}
  \label{fign3}
\end{figure}

%==========================
\begin{figure}[!h]
  \centering
 \includegraphics[height=20cm, width=13cm]{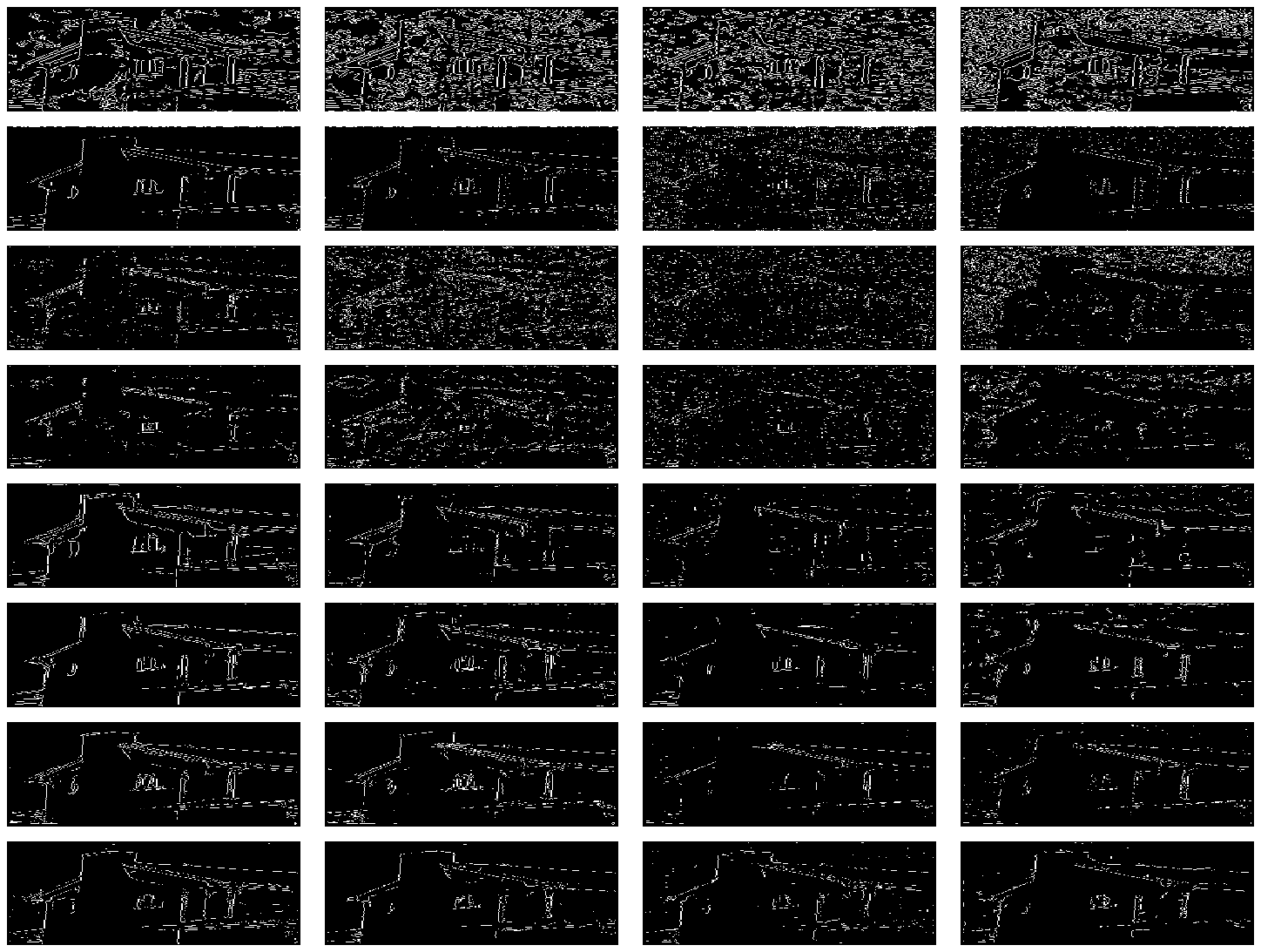}
  \caption{Comparative Results of  Candy, Sobel, DPC, MDPC,  QDLA,MQDLA, SDLA, MSDLA  approaches on  building image from top to the bottom. The first column is added by the  Possion, the second is added by the  Gaussian, the third column is added by the Salt and pepper, the last column is added by the  Speckle noises. }
  \label{fign4}
\end{figure}
\begin{figure}[!h]
  \centering
 \includegraphics[height=20cm, width=13cm]{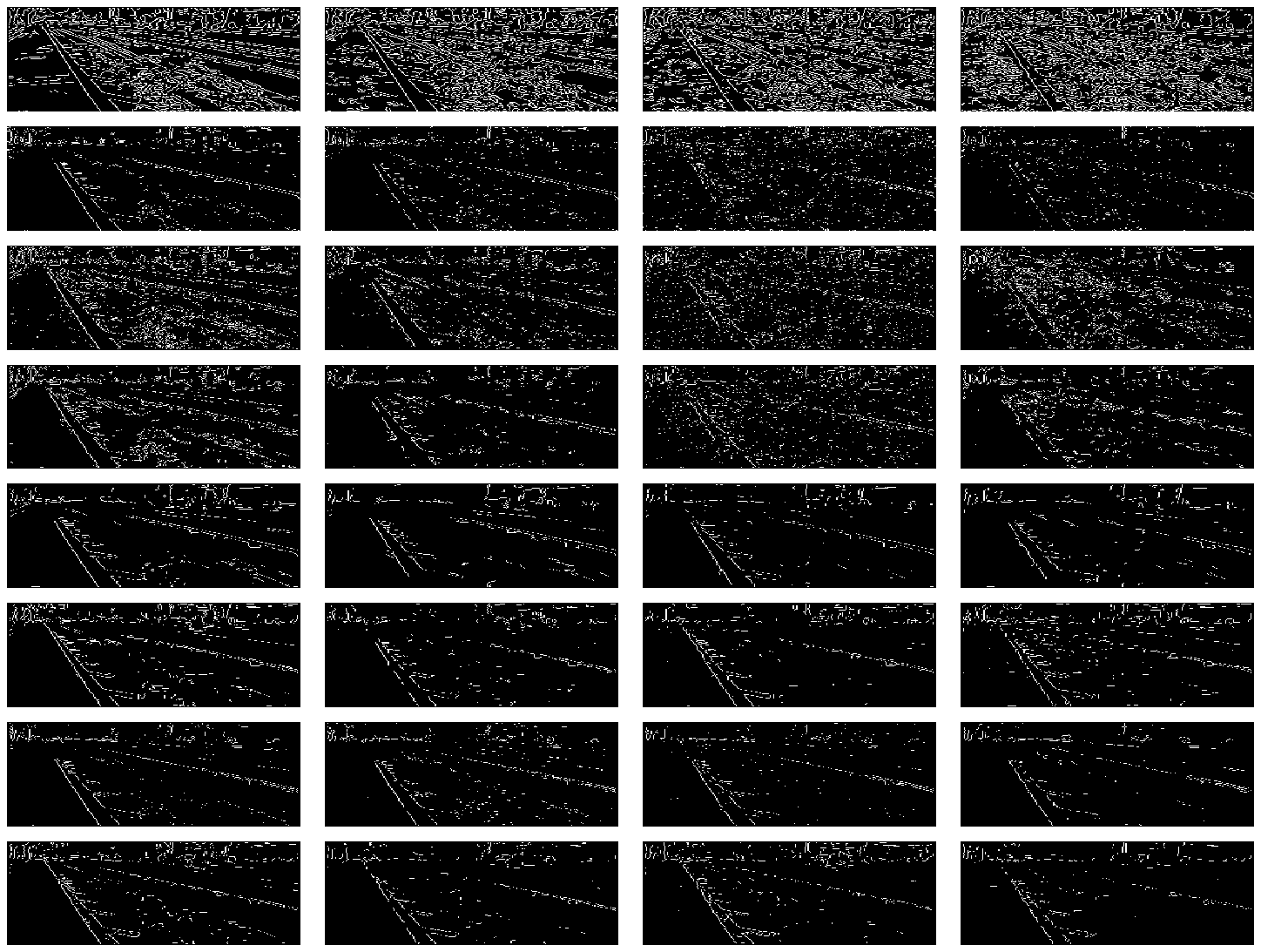}
  \caption{Comparative Results of  Candy, Sobel, DPC, MDPC, QDLA,MQDLA,SDLA, MSDLA  approaches on lane image from top to the bottom. The first column is added by the  Possion, the second is added by the  Gaussian, the third column is added by the Salt and pepper, the last column is added by the  Speckle noises. }
  \label{fign5}
\end{figure}

\begin{figure}[!h]
  \centering
 \includegraphics[height=20cm, width=13cm]{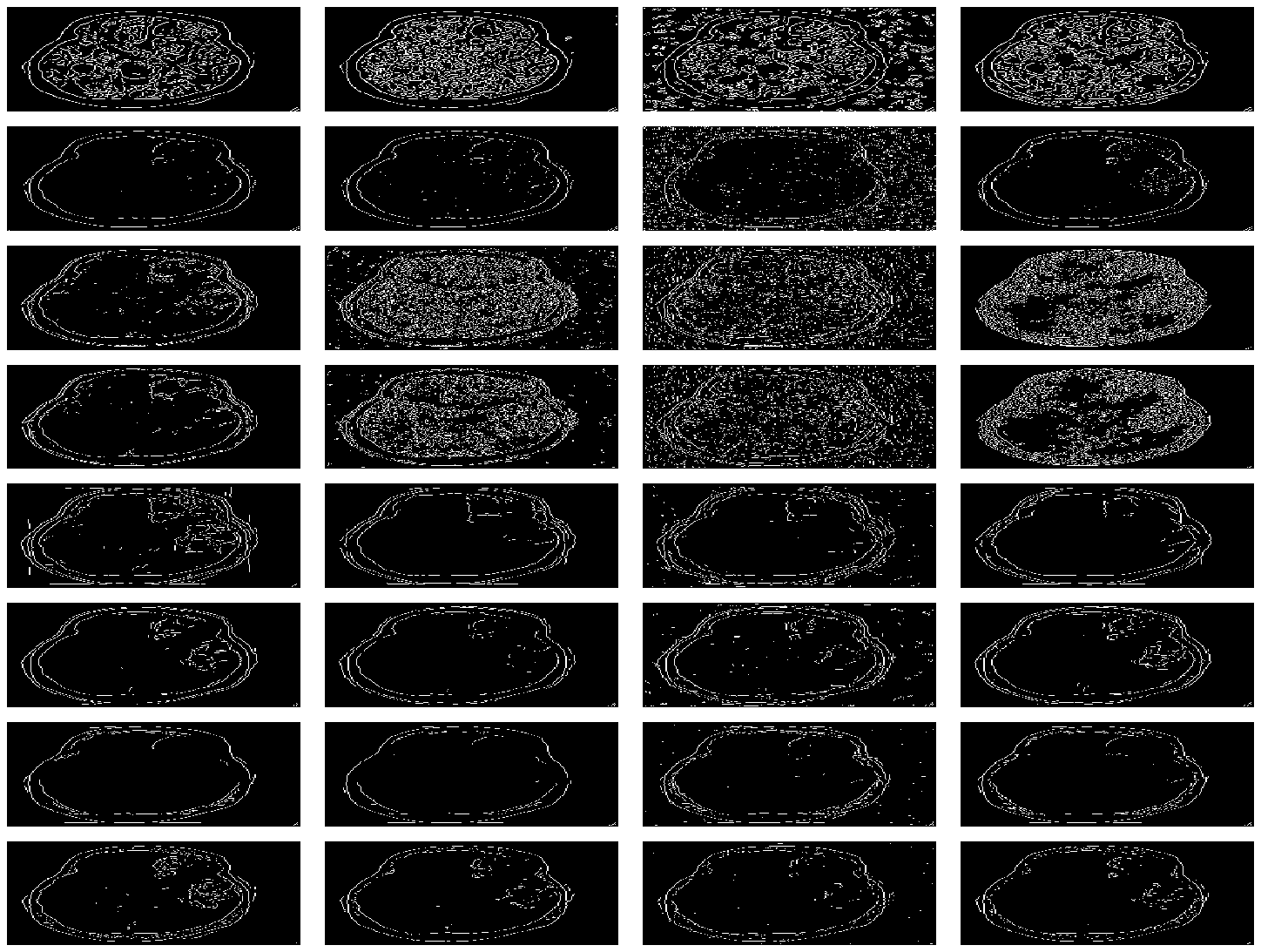}
  \caption{Comparative Results of  Candy, Sobel, DPC, MDPC,  QDLA,MQDLA, SDLA, MSDLA  approaches on  liver image from top to the bottom. The first column is added by the  Possion, the second is added by the  Gaussian, the third column is added by the Salt and pepper, the last column is added by the  Speckle noises. }
  \label{fign6}
\end{figure}

\subsubsection{Quantitative analyses}
\begin{itemize}
\item
In this subsection, two well-known objective image quality metrics, namely the peak-signal-to-noise ratio (PSNR) and the structural similarity index measure (SSIM)\cite{wang2004image}  are analysed.
The \textbf{ PSNR} is a ratio between maximum power of the signal and the power of corrupting noise.
The higher value of PSNR means  the better image quality.
The \textbf{ SSIM} is  correlated with the quality perception of the human visual system.
The SSIM value is between 0 and 1.  0 means no correlation between images, and 1 means two images are exactly the same.
  \item  The PSNR results of various edge detection approaches of the clear image  and different types of noises
  are shown in tables \ref{tab2}, \ref{tab4}, \ref{tab6} and \ref{tab8}.
  The following conclusions are yielded.
      \begin{itemize}
        \item   The MSDLA,  Sobel and  the SDLA are top three approaches.
        \item For the Salt and pepper noise, our proposed approaches  MSDLA, SDLA and  QDLA outperform  than others.
        \end{itemize}

  \item The  SSIM values between the edge detection results  of the clear  image and different types of noises
     are shown in tables  \ref{tab3}, \ref{tab5},  \ref{tab7} and \ref{tab9} show . From  the SSIM  values in these  tables,  we have the following conclusions.
      \begin{itemize}
        \item   From table \ref{tab3}, the three top approaches for house image  are the  MSDLA, the QDLA, the Sobel.
        \item From table \ref{tab5}, the three top approaches for fish image are the MSDLA,  the SDLA, the Sobel.
        \item Form table \ref{tab7}, the three top approaches for lane image are the MSDLA, the MQDLA and the QDLA for Gaussian, Salt and pepper, and Speckle noises. For Possion noise, the top three approaches are Sobel , the MSDLA and the MQDLA.
      \item Form table \ref{tab9}, the three top approaches for liver image are the SDLA, the QDLA, the Sobel.
        \item For the Salt and pepper noise, Our proposed approaches  MSDLA, SDLA, QDLA  outperform than others.
        \end{itemize}

\end{itemize}

%%%%%%%%%%%%%%%%%%%%%%%%%%%%%%%%%%%%%%%%%%%%%%%%%%%%%%%%
\begin{table}[h!]
\caption{PSNR comparison values for the building  image corrupted by different types of noises. }\label{tab2}
\centering
\begin{tabular}{|c|c|c|c|c|}
  \hline

  Building Image &Possion noise  &  Gaussian noise   & Salt and pepper noise & Speckle noise\\
 % \hline
     SNR       &    21.9366   &   15.1953       &    15.1704      & 12.1711 \\
\hline
  QDLA         & 12.8536         &    12.4640      &\textbf{ 12.1802 } &12.0699\\
  \hline
  MQDLA       & 12.3692            & 11.9885          &11.8352        & 11.6404\\
\hline
  SDLA        & \textbf{13.8576}   & \textbf{13.2010} &\textbf{12.7853} &\textbf{13.4110}\\
  \hline
  MSDLA     & \textbf{22.2653}    & \textbf{13.4446} &\textbf{18.6807} &\textbf{13.0943}\\
\hline
  DPC      &  11.3587         & 8.7472               &8.4344         &8.9117\\
 \hline
  MDPC     & 12.2731          & 10.5810              &9.0634          & 10.6015 \\
\hline
 Sobel     &\textbf{ 17.8588}   & \textbf{15.3887}      &11.2699       &\textbf{14.2133}\\
\hline
 Canny    & 10.4607            & 7.6753           &7.7588        &7.2481\\
  \hline
\end{tabular}

\end{table}

%%%%%%%%%%%%%%%%%%%%%%%%%%%%%%%%%%%%%%%%%%%%%%%%%%%%%%%%
\begin{table}[h!]
\caption{SSIM comparison values for the building image corrupted by different types of noises. }\label{tab3}
\centering
\begin{tabular}{|c|c|c|c|c|}
  \hline

  Building Image &Possion noise  &  Gaussian noise   & Salt and pepper noise & Speckle noise\\
  %\hline
     SNR       &  21.0366   &        15.1953      & 15.1704    & 13.1711 \\
\hline
  QDLA      &0.7022        &0.6383             &\textbf{0.5699}     &0.5573\\
  \hline
  MQDLA      & 0.6052              & 0.5512                &  0.4785     & 0.4803\\
\hline
  SDLA      &\textbf{0.7154}      &\textbf{ 0.6952}          &\textbf{0.5591}      &\textbf{0.6212}\\
  \hline
  MSDLA & \textbf{0.9299}      & \textbf{0.6957}      &\textbf{0.7240}       &\textbf{0.6557}\\
\hline
  DPC &   0.4412               & 0.1357            &0.0778               &0.1997 \\
 \hline
  MDPC & 0.5505                  & 0.3541          &0.1036              & 0.3495 \\
\hline
Canny& 0.5133                  & 0.2679                &0.2594                 &0.2645\\
\hline
 Sobel & \textbf{0.8831}         &  \textbf{0.7297}    &0.2528             &\textbf{0.5842}\\
  \hline
\end{tabular}

\end{table}

%%%%%%%%%%%%%%%%%%%%%%%%%%%%%%%%%%%%%%%%%%%%%%%%%%%%%%%%%%%

%%%%%%%%%%%%%%%%%%%%%%%%%%%%%%%%%%%%%%%%%%%%%%%%%%%%%%%%
\begin{table}[h!]
\caption{PSNR comparison values for the fish image corrupted by different types of noises. }\label{tab4}
\centering
\begin{tabular}{|c|c|c|c|c|}
  \hline

  Fish Image &Possion noise  &  Gaussian noise   & Salt and pepper noise & Speckle noise\\
  %\hline
  SNR       &   25.0937   &   18.8141       &    18.8221        & 17.7628 \\
\hline
  QDLA         & 14.5343          &    13.2038       &\textbf{ 13.0903 } &12.4840\\
  \hline
  MQDLA       & 13.5543           & 13.0246          &12.0444        & 13.2878\\
\hline
  SDLA        & \textbf{16.3664}   & \textbf{15.1091} &\textbf{14.6779} &\textbf{14.6604}\\
  \hline
  MSDLA     & \textbf{15.7743}    & \textbf{14.5419} &\textbf{14.2802} &\textbf{14.6199}\\
\hline
  DPC      &  13.6042          & 10.9823              &10.3732        &11.4006\\
 \hline
  MDPC     & 14.2133          & 12.5110               &11.1096         & 12.6723 \\
\hline
 Sobel     &\textbf{ 21.1900}   & \textbf{18.0154}      &11.5467        &\textbf{16.5963}\\
\hline
 Canny    & 12.1900             & 8.7313         &8.8579        &7.3773\\
  \hline
\end{tabular}

\end{table}

%%%%%%%%%%%%%%%%%%%%%%%%%%%%%%%%%%%%%%%%%%%%%%%%%%%%%%%%
\begin{table}[h!]
\caption{SSIM comparison values for the fish image corrupted by different types of noises. }\label{tab5}
\centering
\begin{tabular}{|c|c|c|c|c|}
  \hline

  Fish Image &Possion noise  &  Gaussian noise   & Salt and pepper noise & Speckle noise\\
  %\hline
     SNR      &   25.0937   &   18.8141       &    18.8221        & 17.7628\\
\hline
  QDLA      &0.8035       &0.7365     &\textbf{0.7107}             &0.6848\\
  \hline
  MQDLA      & 0.7522         & 0.7132                &  0.6001      & 0.7125\\
\hline
  SDLA      &\textbf{0.8238}  & \textbf{0.8001}            &\textbf{0.7009}      &\textbf{0.7665}\\
  \hline
  MSDLA & \textbf{0.8131}      & \textbf{0.7569}      &\textbf{0.7038}       &\textbf{0.7529}\\
\hline
  DPC &   0.7396              & 0.4992           &0.1867                &0.5401\\
 \hline
  MDPC & 0.7749                   & 0.6554          &0.2915               & 0.6604 \\
\hline
Canny& 0.6342                   & 0.4562                &0.3292                &0.3106 \\
\hline
 Sobel & \textbf{0.9320}         &  \textbf{0.8285}    &0.3157             &\textbf{0.7463}\\
  \hline
\end{tabular}

\end{table}

%%%%%%%%%%%%%%%%%%%%%%%%%%%%%%%%%%%%%%%%%%%%%%%%%%%%%%%%%%%

%%%%%%%%%%%%%%%%%%%%%%%%%%%%%%%%%%%%%%%%%%%%%%%%%%%%%%%%
\begin{table}[h!]
\caption{PSNR comparison values for the lane image corrupted by different types of noises. }\label{tab6}
\centering
\begin{tabular}{|c|c|c|c|c|}
  \hline

  Lane Image &Possion noise  &  Gaussian noise   & Salt and pepper noise & Speckle noise\\
  %\hline
  SNR       &   22.0935  &   15.2766     &      13.6430   & 22.0774 \\
\hline
  QDLA         & 14.3320        &   13.2512       & \textbf{ 13.6430}  & 12.9391 \\
  \hline
  MQDLA       & 14.9426           &   13.6255      &  13.0074     & 12.9169 \\
\hline
  SDLA        & \textbf{15.6826}   & \textbf{14.2150} & \textbf{ 13.9144} &\textbf{14.1387}\\
  \hline
  MSDLA     & \textbf{15.6777}    & \textbf{14.5740 } &   \textbf{ 13.6209}    &\textbf{13.7620}\\
\hline
  DPC      &  8.0192         & 7.6057             &  7.2626  &7.2760\\
 \hline
  MDPC     & 8.5859         & 8.2340               &    7.8697    & 8.0899 \\
\hline
 Sobel     &\textbf{ 17.7746}   & \textbf{14.7715}      &  12.1496     &\textbf{13.9127}\\
\hline
 Canny    & 11.6715            & 8.3611         &    7.6715   &7.3495 \\
  \hline
\end{tabular}

\end{table}

%%%%%%%%%%%%%%%%%%%%%%%%%%%%%%%%%%%%%%%%%%%%%%%%%%%%%%%%
\begin{table}[h!]
\caption{SSIM comparison values for the lane image corrupted by different types of noises. }\label{tab7}
\centering
\begin{tabular}{|c|c|c|c|c|}
  \hline

  Lane Image &Possion noise  &  Gaussian noise   & Salt and pepper noise & Speckle noise\\
  %\hline
   SNR          &   22.0935  &   15.2766     &      13.6430   & 22.0774 \\
\hline
  QDLA      &0.7217               &\textbf{0.6685}     &\textbf{0.6295}           &\textbf{0.6294}\\
  \hline
  MQDLA      &\textbf{ 0.7618 }    &  \textbf{0.6489 }      & \textbf{ 0.6100}      &0.5989\\
\hline
  SDLA      & 0.7329               & 0.6424           &0.6069               &\textbf{0.6466}\\
  \hline
  MSDLA &\textbf{ 0.7743 }          & \textbf{0.6919}      &\textbf{0.6529}       &\textbf{0.6437}\\
\hline
  DPC &   0.2441                     & 0.1472        &  0.0811                     &0.1301\\
 \hline
  MDPC & 0.2871                     & 0.1882 & 0.1195                       & 0.1859\\
\hline
Canny& 0.7118                        & 0.4349             &0.3675                &0.3387 \\
\hline
 Sobel & \textbf{0.8062}         &  0.6236  &0.3300                              &0.5377\\
  \hline
\end{tabular}

\end{table}

%%%%%%%%%%%%%%%%%%%%%%%%%%%%%%%%%%%%%%%%%%%%%%%%%%%%%%%%%%%
%%%%%%%%%%%%%%%%%%%%%%%%%%%%%%%%%%%%%%%%%%%%%%%%%%%%%%%%
\begin{table}[h!]
\caption{PSNR comparison values for the liver image corrupted by different types of noises. }\label{tab8}
\centering
\begin{tabular}{|c|c|c|c|c|}
  \hline

  Liver  Image &Possion noise  &  Gaussian noise   & Salt and pepper noise & Speckle noise\\
  %\hline
  SNR     & 23.1723  &   14.2158       &    9.9592       & 14.3338\\
\hline
  QDLA      &  13.8431  & 14.6314          &\textbf{ 13.1914 } &14.3338\\
  \hline
  MQDLA      & 13.9111 & 13.9586             &12.4828        & 12.8169\\
\hline
  SDLA       & \textbf{ 15.9428}& \textbf{15.8476}    &\textbf{14.2909} &\textbf{14.7327}\\
  \hline
  MSDLA    &\textbf{15.5237 }& \textbf{16.0723}     &\textbf{15.1567} &\textbf{15.3658}\\
\hline
  DPC    &  11.5193&  8.1548                  &8.7022       &8.4423\\
 \hline
  MDPC    & 12.6735& 9.2078                     &9.1819         & 9.1214\\
\hline
 Sobel    & \textbf{ 23.3218} &\textbf{19.4960 }        &12.3385        &\textbf{17.7509}\\
\hline
 Canny   &12.7378  & 9.8212               &8.6939       &9.9151\\
  \hline
\end{tabular}

\end{table}

%%%%%%%%%%%%%%%%%%%%%%%%%%%%%%%%%%%%%%%%%%%%%%%%%%%%%%%%
\begin{table}[h!]
\caption{SSIM comparison values for the liver image corrupted by different types of noises. }\label{tab9}
\centering
\begin{tabular}{|c|c|c|c|c|}
  \hline

  Liver  Image &Possion noise  &  Gaussian noise   & Salt and pepper noise & Speckle noise\\
  %\hline
   SNR      &  23.1723&    14.2158      &    9.9592       & 14.3338\\
\hline
  QDLA     &  0.7253 &0.7921        &\textbf{0.6924}           &\textbf{0.7936}\\
  \hline
  MQDLA    & 0.7207  & 0.6958             &  0.6820      &0.6661\\
\hline
  SDLA     & \textbf{0.8161} &\textbf{0.8026}  & \textbf{0.7456}      &\textbf{0.7863}      \\
\hline
  MSDLA     &   \textbf{0.8073}& \textbf{0.8075}      & \textbf{0.7439}      &0.7787    \\
\hline
  DPC        &0.5531    &   0.2835                  &0.2178              &0.4564\\
 \hline
  MDPC         & 0.6307  & 0.4184                &0.2814               & 0.4913\\
\hline
Canny           & 0.7760       & 0.6437              &0.3858               &0.6142 \\
\hline
 Sobel            & \textbf{0.9465}    &  \textbf{0.8695}  &0.3638            &\textbf{0.8350}\\
  \hline
\end{tabular}

\end{table}

\section{Conclusion and Future Work} \label{hxx6}
 %The quaternion analytic signal provides the signal features representation, such as the local attenuation and local phase vector.
 In this paper,  firstly, by using the generalized Cauchy$ -$Riemann equations, we not only obtain
the relations between the local features of quaternion Hardy function in higher
dimensional spaces in Corollary  \ref{coro11}, but also  in Theorem \ref{th1} gives the more detail relations
between the  four components of $f$ and its local attenuation in higher
dimensional spaces.
Secondly, there are different  types of edge detection filters, which are connected between components of  the quaternion Hardy function   and its  local features.  Finally, the comparison examples  have numerically confirmed  that  our proposed approaches perform better than the conventional ones.  on one hand, our proposed approaches can not only detect the whole smooth region, but also the  local small change region. On the other hand,  our  approaches have a good performance in  denoising various types of the noised images.
\par
In the future,
%on the one hand, For $q= r +m_1\i + m_2\j + m_3\k \in \mathbb{H},$  we can use b; c and d to represent, respectively, the R, G, and B values of a color image pixel, and set $a = 0$, then  we attempt to apply the  edge detection  approach  on the color image.
%On the other hand,
Kou \cite{kou2016envelope} defined the generalized quaternion analytic signal in invoking  the quaternion linear canonical transform, which has six parameters, when these parameters chosen the special values, the quaternion linear canonical transform reduces to quaternion Fourier transform, and Kou \cite{kou2016envelope} applied the generalized quaternion analytic signal to do envelope of the image, which inspires us to develop more image processing applications  by using the generalized quaternion analytic signal.

%%%%%%%%%%%%%%%%%%%%%%%%%%%%section666666666666%%%%%%%%%%%%%%%%%%%%%%%%%%%%%%%%%%%%%%%%%%%%%%%%%%
\subsection*{Acknowledgments}
Kit Ian Kou acknowledges financial support by The Science and Technology Development Fund, Macau SAR (File no. FDCT/085/2018/A2).%The Fundação para a Ciência e a Tecnologia
Xiaoxiao Hu acknowledges financial support from the Research Development Foundation of Wenzhou Medical University (QTJ18012)

%========================================================

%\bibliographystyle{siam}
%\bibliography{edgedection2}

%%%%%%%%%%%%%%%%%%%%%%%%%%

\end{document}